\documentclass[useAMS,usenatbib,usegraphicx]{mn2e}

\usepackage{amsmath}
\usepackage{amssymb}
\usepackage{xspace}

\newcommand{\etal}{et~al.\xspace}
\newcommand{\etc}{etc.\@\xspace}
\newcommand{\ie}{i.e.,\xspace}
\newcommand{\eg}{e.g.,\xspace}

\newcommand{\wmap}{\emph{WMAP}\xspace}

\def\ltsim{\ifmmode\stackrel{<}{_{\sim}}\else$\stackrel{<}{_{\sim}}$\fi}
\def\gtsim{\ifmmode\stackrel{>}{_{\sim}}\else$\stackrel{>}{_{\sim}}$\fi}
\newcommand{\mbold}[1]{\mbox{\boldmath${#1}$}}

\newcommand{\beq}{\begin{equation}}
\newcommand{\eeq}{\end{equation}}

\title[Modelling the Galactic Magnetic Field]{Modelling the Galactic Magnetic Field on the Plane in 2D}
\author[T. R. Jaffe et al.]{T. R. Jaffe $^{1}$\thanks{E-mail:tess.jaffe@manchester.ac.uk}, 
   J. P. Leahy$^{1}$\thanks{E-mail:j.p.leahy@manchester.ac.uk}, 
   A. J. Banday$^{2,3}$\thanks{E-mail:Anthony.Banday@cesr.fr}, 
   S. M. Leach$^{4,5}$\thanks{E-mail:leach@sissa.it},\newauthor
   S. R. Lowe$^{1}$\thanks{E-mail:Stuart.Lowe@manchester.ac.uk}, 
   A. Wilkinson$^{1}$\thanks{E-mail:althea.wilkinson@manchester.ac.uk}\\
$^{1}$Jodrell Bank Centre for Astrophysics, School of Physics and Astronomy, The University of Manchester, Oxford Road, \\
\hspace{1cm} Manchester, M13 9PL, United Kingdom\\
$^{2}$Centre d'Etude Spatiale des Rayonnements, 9 av. du Colonel Roche, BP~44346, 31028 Toulouse Cedex 5, France\\
$^{3}$Max Planck Institute for Astrophysics, Karl-Schwarzschild Str. 1, 85741 Garching, Germany\\
$^4$SISSA, Astrophysics Sector, via Beirut 2-4, I-34014 Trieste, Italy.\\
$^5$INFN, Sezione di Trieste, I-34014 Trieste, Italy.\\
}
\begin{document}

\date{}

\pagerange{\pageref{firstpage}--\pageref{lastpage}} \pubyear{}

\maketitle

%%%%%%%%%%%%%%%%%%%%%%%%%%%%%%%%%%%%%%%%%%%%%%%%%%%%%%
\label{firstpage}

\begin{abstract}
We present a method for parametric modelling of the physical
components of the Galaxy's magnetised interstellar medium,
simulating the observables, and mapping out the likelihood space using
a Markov Chain Monte-Carlo analysis.  We then demonstrate it
using total and polarised synchrotron emission data as well as
rotation measures of extragalactic sources.  With these three
datasets, we define and study three components of the magnetic field:
the large-scale coherent field, the small-scale isotropic random
field, and the ordered field.  In this first paper, we use only data
along the Galactic plane and test a simple 2D logarithmic spiral model
for the magnetic field that includes a compression and a shearing of
the random component giving rise to an ordered component.  We
demonstrate with simulations that the method can indeed constrain
multiple parameters yielding measures of, for example, the ratios of
the magnetic field components.  Though subject to uncertainties in
thermal and cosmic ray electron densities and depending on our
particular model parametrisation, our preliminary analysis shows that
the coherent component is a small fraction of the total magnetic field
and that an ordered component comparable in strength to the isotropic
random component is required to explain the polarisation fraction of
synchrotron emission.  We outline further work to extend this type of
analysis to study the magnetic spiral arm structure, the details of
the turbulence as well as the 3D structure of the magnetic field.
\end{abstract}


%%%%%%%%%%%%%%%%%%%%%%%%%%%%%%%%%%%%%%%%%%%%%%%%%%%%%%
\section{Introduction}\label{intro}
%%%%%%%%%%%%%%%%%%%%%%%%%%%%%%%%%%%%%%%%%%%%%%%%%%%%%%

Observations of external galaxies show that the relationship between
the magnetic field and the ionised gas is far from simple and varies from
galaxy to galaxy.  (See, \eg \citealt{beck:2009} for a review.)  Our
own galaxy is more difficult to study since we must observe it from
within and looking through the plane.  In this work, we present a
method for comparing magnetic field models of many parameters to a
variety of observables.  With the tools we describe, we can use more
realistic models for the small-scale components and therefore have
less need for the simplifying assumptions (such as isotropy) that are
often made with unknown effect on the results.  For the first time, we
now have coverage of significant portions of the Galactic plane in
total and polarised synchrotron intensity as well as in rotation
measures (RM), and we show how these complementary datasets are vital
for disentangling the different components of the Galactic magnetic
field.

\begin{figure}
\includegraphics[width=\linewidth]{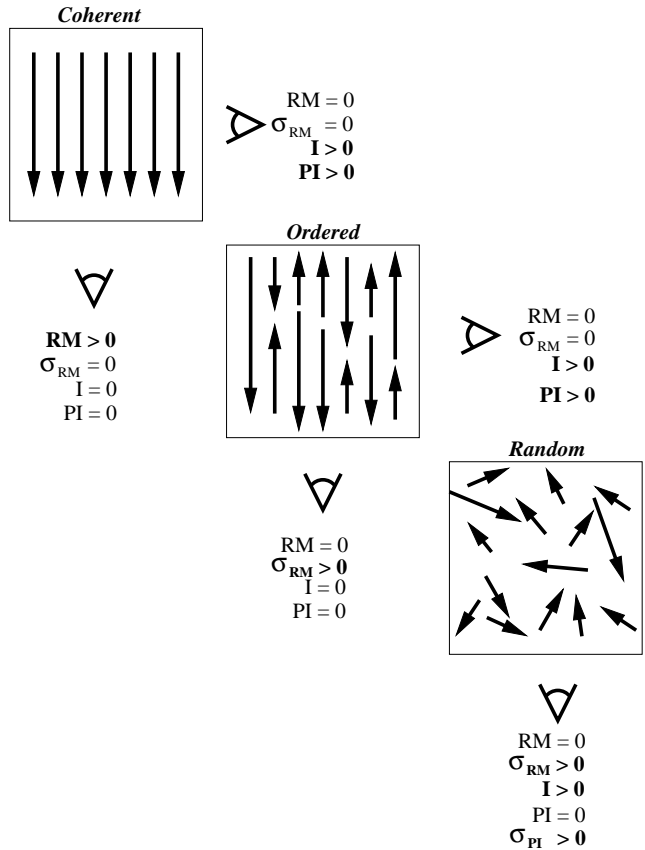}
\caption{Cartoon illustrating the three components of the magnetic field 
and how they relate to the three observables of total synchrotron
intensity (I), polarised synchrotron intensity (PI), and Faraday
rotation measure (RM).  
%%%%%%%%% TO BE REMOVED:  BOLD
{\bf 
The variance of these is also of interest, as
shown in the case of $\sigma_\rmn{RM}$.  
}
%%%%%%%%%%%
(Note that the situation is
the same for dust emission, which is also polarised perpendicular to
the magnetic field due to grain alignment.)}
\label{fig:cartoon}
\end{figure}

It is useful to think of the Galactic magnetic field as separable into
three components referred to as coherent, ordered, and
random/tangled/turbulent.  These are illustrated by the cartoon in
Fig.~\ref{fig:cartoon}.  The term ``coherent'' refers to, \eg a
large-scale spiral structure, while the ``random'' component usually
refers to the small-scale component varying in three dimensions in
both strength and direction.  The term ``ordered'' refers to a field
where variations simply imply sign reversals; the ordered component
points along a common, ordered axis but simply changes direction on
small scales, perhaps stochastically.  (There are different uses of
the word ``ordered'' in the literature.  It can be thought of as an
anisotropic random component, but in this paper, we will distinguish
between an isotropic random component and an ordered component.
``Ordered'' is also sometimes used to refer to the combination of what
we in this paper call the ordered plus coherent field, but we find that
usage confusing and prefer to think of them as three distinct
components.)

The coherent fields are assumed to probe processes such
as galactic-scale dynamos, while the small-scale random component
probes turbulent processes in the ISM.  The ordered component is
thought to result from the effects of larger scale shearing dynamics,
compression, \etc on the random component.  This component is often
neglected for simplicity (see, \eg \citealt{burn:1966}).

In external galaxies, the synchrotron emission is compared to the
thermal emission to study how the magnetic fields correlate with the
ionised gas in the galaxy.  The arms seen in total intensity tend to
follow the arms seen in thermal emission, but the same is not true of
the polarised intensity.  This has two implications.  Firstly, the
random component is often significant, perhaps dominant, in the spiral
arms, thus lowering the polarisation fraction.  Secondly, the ordered
and coherent fields may be strongest in ``magnetic arms'' distinct
from the spiral arms traced by the diffuse ionised gas (DIG)
component.  As reviewed by \citet{beck:2009}, NGC~6946 is an example
where the magnetic arms appear to be between the arms traced by
thermal emission, while M~51 appears to have magnetic arms on the edge
of the spiral arms.

\begin{figure}
%
%  work/galactic_b/plane_mcmc/try_data2/plots/plane_annotated.eps
%  IDL> plot_planes3,[''],/nosims,/anno,psfile='../plots/plane_annotated.eps'
%
\includegraphics[width=\linewidth]{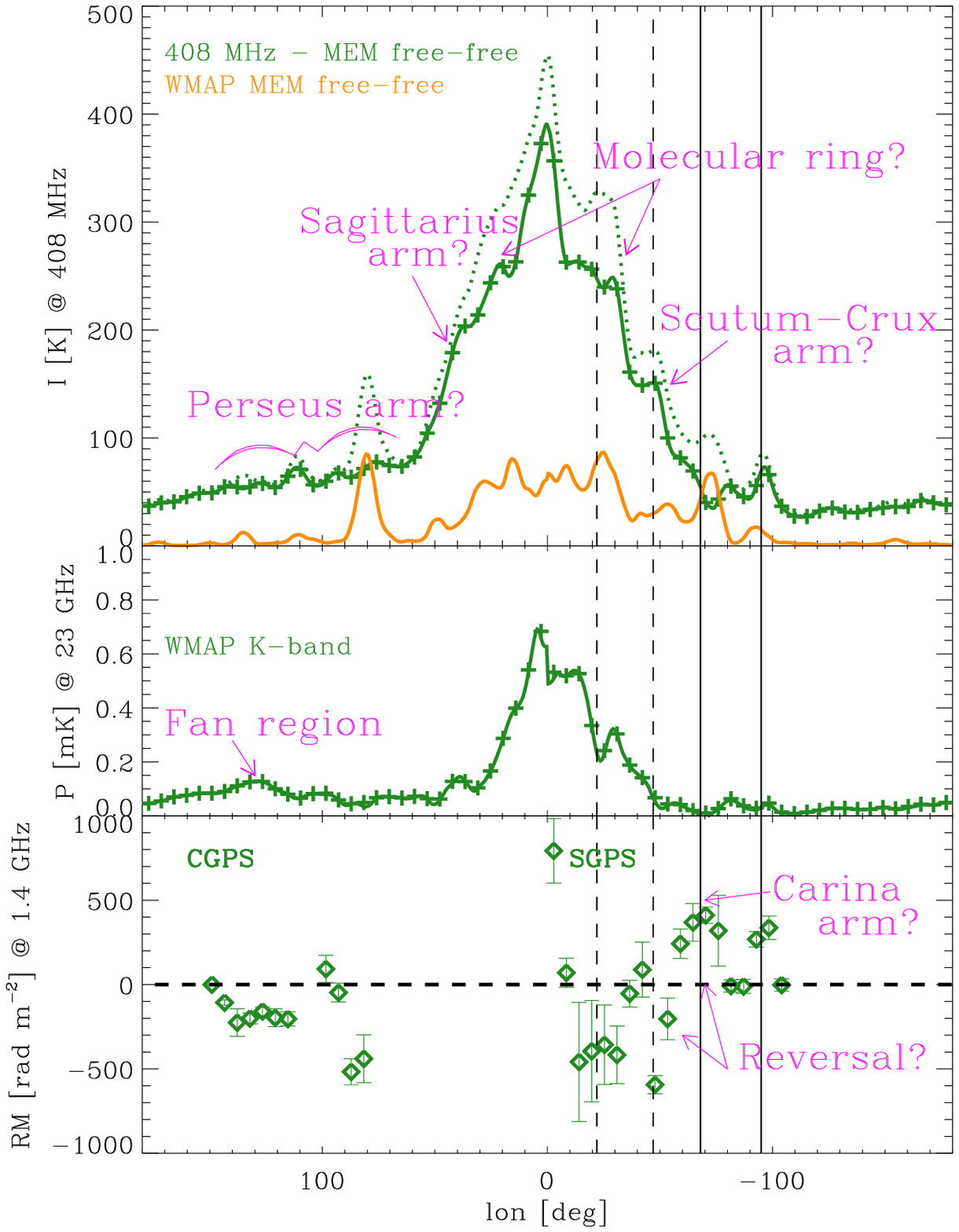} 
\caption{The three primary tools we use for studying the magnetic field in the 
plane of the galaxy are the synchrotron total intensity ({\em top})
from \citet{haslam:1982}, the synchrotron polarised intensity ({\em
middle}) from \citet{hinshaw:2009}, and the RM data ({\em bottom})
from \citet{brown:2003,brown:2007} 
%%%%% TO BE REMOVED:  BOLD
{\bf averaged into roughly $6\degr$ bins}.
%%%%%%%%%
The
field structures that we can perhaps infer from the profile of these
data along the plane are indicated.  The vertical lines mark
interesting sight-lines
%%%%%%%% TO BE REMOVED:  BOLD
{\bf 
(solid for positive RM, dashed for negative)}, 
%%%%%%%
also
shown on Fig.~\ref{fig:regmodel}, showing what may be tangents
to arm features.  Also
shown in orange is the estimate for the free-free contamination which
has been subtracted from the total emission at 408~MHz (dotted green
line); see
\S~\ref{sec:synch}.}
\label{fig:plane_dataonly}
\end{figure}

Though we cannot assume anything about how the magnetic field
components correlate with the DIG, Fig.~\ref{fig:plane_dataonly}
shows that the synchrotron emission has clear step features that have
been thought to correspond to arm tangents (\eg \citealt{mills:1959}
or \citealt{beuermann:1985}).  The rotation measures also indicate field
reversals that may be related.  Learning how the field varies in
strength, direction, and coherence across a spiral arm traced by
diffuse gas will inform theories of how the fields are generated and
maintained in the dynamic environment of the magnetised interstellar
medium (MIM).

In this work, we take a first step toward disentangling these
components in our own Galaxy.  Since we cannot look down upon it from
above to determine where the magnetic arms lie relative to the DIG 
spiral arms, we are limited to looking through the plane.  Our ongoing
project is to see whether the profiles in total and polarised
synchrotron emission can distinguish arm ridges in the different
components, but for this work, we assume that they all peak in the
same ridges.  We do not, however, constrain those ridges to lie where
the DIG spiral arm ridges are thought to be, but instead use the
data to constrain the orientation of the magnetic spiral arms.  We
leave it to later work to determine if the data can distinguish models
where in addition to independent magnetic arms, the peaks in the three
components do not coincide.

%If the variance of the random component increases in the
%arms along with the amplitude of the coherent component (maintaining a
%constant ratio), we will then see increased emission when looking
%along the arms.  But \cite{brindle:1978} found that such a simple
%amplitude enhancement produced peaks in the profile that were too
%spikey to match the observations.  This problem is naturally resolved
%if, in addition to the amplitude enhancement, the random component is
%also anisotropic in the arms, namely elongated in the direction of the
%spiral.  Equivalently, there is an ordered component of the field in
%the arms.  The physical motivation for such anisotropy is the shearing
%effect that the shock front in the arm might have on the magnetic
%field.

But there are several difficulties in interpreting the emission
profile of synchrotron total intensity along the Galactic plane.
Firstly, the random component introduces a sort of ``galactic
variance'' (analogous in a limited sense to the cosmic variance in
cosmic microwave background studies).  Nearby random features will
perturb the profile making it difficult to separate the field
components accurately.  Secondly, the plane contains a strong emission
component from thermal bremsstrahlung, commonly known as free-free
emission.  This component is difficult to separate from the
non-thermal emission we wish to study and has a different spatial
distribution.

Using the polarised synchrotron emission in addition helps to resolve
some of these problems, as the isotropic random component does not on
average contribute to the polarised intensity.  The addition of
rotation measures further helps to break the degeneracies in the
parameter space, as they are only dependent on the coherent component.
The three field components can then be studied by comparing these
three observables.

It is only recently that we have enough data to study the three
components of the magnetic field and perhaps to separate them.  We can
look at the continuum emission and compare the total intensity to
polarised intensity as well to the Faraday rotation measures.  These
three contain information about the relative strengths of the
coherent, ordered, and random components.  We now have synchrotron
emission maps over the full sky at several bands including
polarisation, and we have unprecedented coverage of a large part of
the Galactic plane in RMs of extragalactic sources.  The combination
of these datasets allows us to probe the three-dimensional
distribution of magnetic fields and ionised gas in the Galaxy.

There has been much previous work to model the Galactic magnetic field,
from \citet{beuermann:1985} and \cite{broadbent:1990} to
\citet{han:2006}, \citet{page:2007}, \citet{deschenes:2008}, 
\citet{sun:2008}, \citet{jansson:2009}, and \citet{orlando:2009}.  
Most of these works use only one or two datasets and are limited by
simplifying assumptions about the nature of the irregular components.
\citet{broadbent:1990} (with more details in \citealt{broadbent:1989})
consider a detailed galaxy model and have remarkable success in
reproducing the profile of synchrotron emission along the plane.  That
work is unusual in including an analytic method to model the
anisotropy in the small-scale random component of the MIM.  While it
seems unlikely that the data constrain all of the dozens of parameters
in their model (particularly considering the above-mentioned problems
of thermal emission separation and galactic variance), we consider it
an interesting place to start and adopt several aspects of their
model.  The more recent work of \citet{sun:2008} uses the three
observables together to present a 3D model of the galaxy, but it does
not explore the parameter space systematically and considers only what
we define as the coherent and isotropic random magnetic field
components.  The work of \citet{jansson:2009} does a systematic search
using MCMC, but they do not explicitly treat the random (or ordered)
components and use only two of the three datasets we consider vital.

Our principle aim in this work is to describe a method to model the
physical components of the MIM, to simulate the observables, and to
map out the likelihood space using a Markov Chain Monte-Carlo (MCMC)
analysis.  The problem is a large and complicated one, and in this
first paper, we focus on a 2-dimensional analysis on the Galactic
plane only (seen as a 1D profile as in Fig.~\ref{fig:plane_dataonly}),
though the methods can certainly be extended to model the Galaxy in
3D.  We then demonstrate the method's utility by using the available
data to constrain the relative strengths of the three components of
the magnetic field, the coherent, random, and ordered fields, and
discuss how to extend the analysis using additional data.

%%%%%%%%%%%%%%%%%%%%%%%%%%%%%%%%%%%%%%%%%%%%%%%%%%%%%%%%
\section{Observations}
%%%%%%%%%%%%%%%%%%%%%%%%%%%%%%%%%%%%%%%%%%%%%%%%%%%%%%%%

%%%%%%%%%%%%%%%
\subsection{\bf Synchrotron emission in total intensity} \label{sec:synch}

The 408~MHz full-sky map by \cite{haslam:1982} primarily consists of
synchrotron emission.  (This map is available in
HEALPix\footnote{\tt{http://healpix.jpl.nasa.gov/}} format from the
LAMBDA website\footnote{\tt{http://lambda.gsfc.nasa.gov/}}.)  The
intensity gives information about the magnetic field ($B$) component
perpendicular to the line of sight:
\beq \label{eq:synch}
I_\rmn{sync}(\nu,p) \propto \int_\rmn{LOS} J_\rmn{CRE}(\mathbf{x}) B_\perp(\mathbf{x})^{\frac{p+1}{2}} \rmn{d}l \eeq
where $I$ is the specific intensity and $J_\rmn{CRE}(\mathbf{x})$ is
the density of cosmic ray electrons (CREs) described explicitly in
\S~\ref{sec:cres}.  The total intensity, or Stokes I, in a given 
observing beam is then I$=\int Id\Omega$.

The index on $B$ depends on the spectral index, $p$, of the cosmic ray
electron distribution, which we assume to be a power law.  We follow
the notation of \citet{rybicki:1979} and define the number density of
particles in the range $\gamma$ to $\gamma +d\gamma$ (where $\gamma$
is the Lorentz factor such that $E=\gamma m_\rmn{e}c^2$) as
$N(\gamma)d\gamma\propto\gamma^{-p}d\gamma$.  Observations show that
$p\approx 3$ between our two synchrotron observing frequencies of
408~MHz and 23~GHz (see, \eg \citealt{finkbeiner:2004} or
\citealt{deschenes:2008}), in which case $dI\propto B^2$.  It also
implies a power law in synchrotron flux density of $S\propto
\nu^{-(p-1)/2}=\nu^{-1}$ or brightness temperature of $T\propto
\nu^{-3}$.  In reality, the spectral energy distribution of electrons will 
vary with position in the galaxy, but we will approximate it as a
power law with a constant $p=3$.  The impact of this choice is
discussed further in \S~\ref{sec:disc}.

The synchrotron total intensity is sensitive to all components of the
magnetic field.  The coherent component, random component, and ordered
component all contribute according to their projection perpendicular
to the line of sight.  The profile of the synchrotron emission along
the Galactic plane, shown in Fig.~\ref{fig:plane_dataonly}, has
distinct steps, exactly as if the emission increases as the observer
looks along a spiral arm.  Unless this is coincidental, it therefore
seems that either the magnetic field structure or the CRE density must
follow to some degree the DIG spiral structure.

%But it is particularly interesting to note that the increase is not 
%accounted for by the {\bf coherent} component of the magnetic field.
%In all cases where such a coherent field is observed, it points along
%the spiral modulation, and therefore there is no component
%perpendicular to the line of sight when looking down the spiral.  The
%steps must then be the result of the modulation within the arm of the
%random component.  

The 408~MHz map includes thermal bremsstrahlung (free-free) emission
in a very narrow region about the Galactic plane.  (Other available 
radio surveys include the 2.3~GHz data of
\citealt{jonas:1998} or the 1.4~GHz survey of \citealt{reich:1982}
and \citealt{reich:1986}.  Higher frequencies, however, will have more
free-free contamination due to its flatter spectral index.)  Since it
is along the plane that we are interested in the synchrotron emission,
we must subtract the thermal emission.  We use the Wilkinson Microwave
Anisotropy Probe (\wmap) foreground maps generated from the maximum
entropy method (MEM) described in
\citet{hinshaw:2007}.  The \wmap MEM free-free map for the Ka-band (33~GHz)
contains the best estimate of the free-free distribution that we have
for the Galactic plane region.  (\citep{doc:2008}, for example,
produce a synchrotron map using a combination of 11 different
frequencies, but their principal components analysis results in the
408~MHz profile along the plane being virtually unchanged, indicating
that the free-free component has not been subtracted.)  The MEM maps
are also available from the LAMBDA website.

The data are generated by starting with the full sky HEALPix maps at
$N_\rmn{side}=512$ (pixel size of $6.9$~arcmin) for the Haslam and the \wmap
Ka-band MEM free-free component.  Each map is then smoothed
to an effective beamwidth of 3$\degr$ FWHM and the maps downgraded to
$N_\rmn{side}=128$ (pixel size of $27.5$~arcmin).  From the Haslam data, we
then subtract the free-free component extrapolated from the Ka-band
assuming a power law dependence of $\nu^{-2.1}$ (see, \eg
\citealt{dickinson:2003}).  The 512-pixel ($4N_\rmn{side}$) slice
along the plane is then extracted, further smoothed in 1-dimension by
boxcar-averaging by 8 pixels, and the result is then downgraded to the
64 bins whose centres coincide with the $N_\rmn{side}=16$ pixels
along the plane.  The result is a profile with a resolution of
$\approx6\degr$ in longitude and three in latitude.

The reason for this processing is to minimize the effects of the
inaccurate thermal emission separation.  The \wmap MEM free-free
component is generated by an analysis including only three
foregrounds, namely synchrotron, free-free, and thermal dust.  The
results are therefore contaminated by the anomalous dust component.
There is no other reliable indication of the thermal emission on the
plane, however, (H$\alpha$, for example, is absorbed by dust on the
plane and is only a reliable tracer of free-free at high latitudes) and
the free-free emission is strong enough on the plane relative to the
other components that the separation, while imperfect, is probably
sufficient.  The free-free latitude profile is very narrow (see \eg
\citealt{dickinson:2003}) compared to that of the synchrotron emission.  
Smoothing the maps first then leaves the profile of synchrotron
emission along the plane essentially unchanged but reduces the
relative amount of free-free emission.

In future work, we will investigate other ways to separate the thermal
emission on the plane, for example by using the newer results from
\wmap in \citet{gold:2009}, which include a pixel-based MCMC separation 
method.  In that work, the authors estimate that the spinning dust
fraction is at most $\approx 15-20$ per cent of the emission in the Ka-band.
It is hard to predict the fraction that ends up in the free-free
template (in either the MCMC or MEM analyses) when the spinning dust
component is ignored, since this depends on whether its spectral
behaviour better matches the free-free or the synchrotron.  Note that
the remaining uncertainties due to the inaccurate thermal emission
separation and subtraction are unlikely to be as significant as the
galactic variance described in \S~\ref{intro} itself.

Because the Galactic centre region is likely complicated and currently
mysterious, we exclude the four nearest pixels corresponding to
roughly 10$\degr$ either side of $\ell=0$.  This makes the analysis
insensitive to what happens in the innermost $\sim$2~kpc of the galaxy and
allows us to focus on the step-features as well as the general
profile.

%%%%%%%%%%%%%%%
\subsection{\bf Polarised intensity} 

Polarised synchrotron emission depends on the cosmic ray power law
spectral index, and is at most a fraction $\Pi$ of the total intensity, where 
\beq
\Pi\equiv \frac{PI}{I}=\frac{p+1}{p+7/3}=0.75
\eeq
(for $p=3$) in the case that the magnetic field is
uniform \citep{rybicki:1979}.  The observed degree of polarisation
compared to the total intensity therefore gives information about how
ordered the magnetic field is perpendicular to the line of sight.
Averaging over an isotropic random component produces, on average, no
polarised intensity, as emission polarised in perpendicular directions
cancels.  An {\it ordered} or anisotropic random component, however,
will still add to the polarised intensity, since the polarisation
angle depends only on the orientation rather than the direction of the
field.

We use the \citet{hinshaw:2009} five-year \wmap 23~GHz K-band map of
polarised intensity and smooth it to extract the plane as described
above for the Haslam data.  Likewise, we ignore the four pixels toward
the Galactic centre.  This dataset is currently the best polarised
synchrotron map in the high frequency regime where Faraday effects are
negligible.

%%%%%%%%%%%%%%%
\subsection {\bf Faraday Rotation measure} 

The polarisation angle of an electromagnetic wave rotates when
propagating through a magnetised plasma.  The rotation angle changes
as $RM\lambda^2$, where the Faraday rotation measure (RM) is defined
as the the line-of-sight (LOS) integral from a source at a distance
$D$:  
\beq\label{rm}
RM\propto\int_D^0 n_e B_\parallel \rmn{d}l
\eeq
%and 
%\beq
%\psi = RM\lambda^2 + \psi_0
%\eeq
%where $\psi$ is the polarisation angle on the sky clockwise relative
%to the Galactic meridian.  (For synchrotron
%emission, $\psi_0$ is the angle perpendicular to the component of the
%magnetic field projected onto the sky.)  
where $n_e$ is the number density of thermal electrons.  A positive RM
means that $B_\parallel$ points toward the observer.

The RM can be measured for both pulsars within the galaxy as well as
external sources.  These data complement the synchrotron emission,
since the RM is sensitive to direction as well as orientation and to
the parallel component rather than the perpendicular.  Therefore,
averaged RMs trace only the coherent component of the magnetic field.
If the thermal electron density is uncorrelated with the magnetic
field, then any random component or even ordered component will have a
null effect on average, though it will naturally give rise to a
variance in the manner of a random walk.  This means that these data
can help break degeneracies that arise in modelling the synchrotron
emission due to the coherent versus random field parameters.

We are therefore probing three components of the magnetic field using
three main datasets.  There is a further source of uncertainty,
however, namely the distribution of thermal electrons in the galaxy,
discussed in \S~\ref{sec:ne}.  As mentioned above, the relationship
between the DIG and the magnetic field structure is unclear.

The data we use are the RMs for the extragalactic sources in the
Canadian Galactic Plane Survey (CGPS) of \citet{brown:2003} and in the
Southern Galactic Plane Survey (SGPS) of \citet{brown:2007}.  In the
CGPS, there are a total of 380 sources within 5$\degr$ of the plane in
the range $82\degr \leq \ell \leq 146\degr$.  In the SGPS, there are a
total of 148 sources within $1.5\degr$ of the plane in the range
$253\degr \leq \ell \leq 356\degr$.  Following Brown \etal, we exclude
from our analysis the 30 sources in the region $270\degr
\lesssim \ell \lesssim 280\degr$, where there is evidence for 
anomalously low RMs due to a local feature.  To maintain consistency
with the other datasets, we only use the 151 CGPS sources within
1.5$\degr$ of the plane.  That leaves a total of
%118 (SGPS) + 151 (CGPS)
269 sources used in our analysis covering a bit less than half of the
Galactic plane.  The sources are then averaged into the bins
corresponding to those used for the other datasets.

%%%%%%%% TO BE REMOVED:  BOLD
{\bf There are additional RM data for galactic pulsars that may in
future be added to our analysis.  Because each pulsar is at a
different LOS distance through the galaxy model, to include them would
complicate the analysis and add considerable processing time.  For
this work, then, we use only extragalactic sources that give a full
LOS through the galaxy.  }
%%%%%%%%%

%%%%%%%%%%%%%%%%%%%%%%%%%%%%%%%%%%%%%%%%%%%%%%%%%%%%%%%%
\section{Galaxy Models}\label{sec:models}
%%%%%%%%%%%%%%%%%%%%%%%%%%%%%%%%%%%%%%%%%%%%%%%%%%%%%%%%

We must model each of the relevant components of the MIM, namely the
magnetic field (coherent, random, and ordered), the thermal electron
spatial distribution, and the cosmic ray electron spatial and spectral
distribution.

%%%%%%%%%%%%%%%
\subsection{\bf Coherent magnetic field}\label{sec:reg_prof}

We model the coherent magnetic field beginning with a simple
axisymmetric spiral that defines the direction of the
field.  In Sun-centric coordinates, then, the field direction is
\beq
\hat{\bf B} = \sin(\theta_p+\phi-\ell)\hat{\bf x} \notag - \cos(\theta_p+\phi-\ell)\hat{\bf y} 
\eeq
where $\phi$ is the azimuthal angle in the polar coordinate system
with the Sun at the origin, while $\ell$ is the Galactic longitude.
We are working only in the plane, and the vertical component is always
zero.  (We use this form since we will integrate along the LOS from
the Sun as the origin.)  

This defines a spiral field with a pitch angle of $\theta_p=-11.5$.
This angle is roughly that used in the thermal electron density model
described in \S~\ref{sec:ne}. 
%%%%%  TO BE REMOVED:  BOLD
{\bf 
Estimates of $\theta_p$ in the literature vary, however.
\citet{han:1994} get $\theta_p = -8\degr.2 \pm 0\degr.5$ based on RM
surveys, in rough agreement with the \citet{heiles:1996} result based
on starlight polarisation, $\theta_p = -7\degr.2 \pm 4\degr.1$.
\citet{deschenes:2008} find a value of $-8.5\degr$ based on \wmap
polarised synchrotron emission, while \citet{page:2007} and
\citet{jansson:2009} find best-fit pitch angles of as much as
$35\degr$. Jansson et al., however point out that this is highly
model-dependent.  In later work, we will allow this parameter to vary
to see if it can be constrained better, and to investigate whether the
magnetic spiral arms follow the matter spiral arms.
}
%%%%%

\citet{broadbent:1990} defined the coherent field amplitude as:
\beq
B(r)=B_0(1-\exp(-r^2/R_2^2))(\exp(-r^2/R_0^2)+\exp(-r^4/R_1^4))
\eeq
where $r$ is the Galacto-centric radius, and the three scale radii
define the shape.  Outside of the galactic centre region, this is a
simple exponential drop as used in \citet{han:2006} to fit RM data or in 
\citet{deschenes:2008} for polarised emission.  Because of its success 
predicting the synchrotron total intensity profile, even in the inner
galaxy, in \citet{broadbent:1990}, we start with this model.  But we
note firstly that \citet{deschenes:2008} find a constant amplitude
fits the \wmap data better, and secondly, this profile is degenerate
with the cosmic ray density profile described below in
\S~\ref{sec:cres}.  (In fact, Broadbent \etal use a constant CRE density, 
so their profile was essentially accounting for any radial variations
in both components.)

Constraining this profile is beyond the scope of this paper, as it
will require an additional observable such as dust emission to break
the degeneracy.  The default values we use are given in
Table~\ref{tab:params}, though as described in \S~\ref{sec:synch}, we
exclude the innermost longitudes from the analysis, which means the
analysis is not sensitive to $R_1$ and $R_2$.  The value for $R_0$ is
degenerate with the $h_r$ parameter controlling the cosmic ray profile
defined in \S~\ref{sec:cres}, also an exponential disc.  The default
values in Table~\ref{tab:params} were set by eye to approximate the
synchrotron profile so that we could explore the parameters
determining the random versus ordered components.  In a later paper,
we will explore the radial profile parameters in more detail.

\begin{figure}
\includegraphics[width=\linewidth]{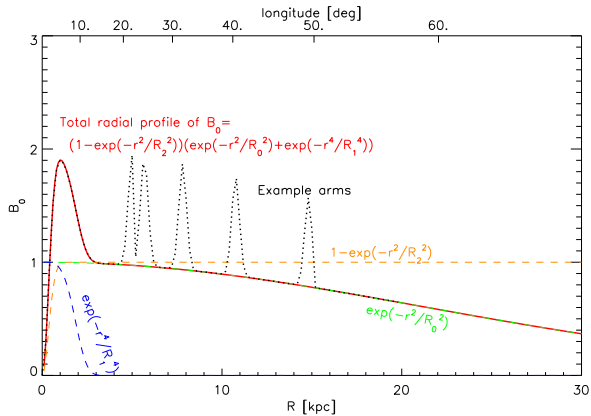}
\caption{An example of the radial profile of the coherent field and of
the compression.  (Note that the values used are just for illustration
and not those found to fit the data.)  The
uncompressed profile is shown in red, while the dotted black line
includes the compression factor, $\rho_c(d)$, showing example spiral
arms.  The bottom axis is the radius from the Galactic centre, while
the top axis is the corresponding longitude if viewed from the
perpendicular direction at a distance of $R_\oplus=8.5$~kpc. }
\label{fig:r_prof}
\end{figure}

This azimuthally symmetric field is then amplified along spiral arm
ridges (defined parallel to the field direction) as described in
\S~\ref{sec:compression}.  Note that those ridges are not constrained
to lie along the DIG spiral arm ridges, and each magnetic arm has a
corresponding amplitude parameter, $a_n$, so that we can fit for the
strength of each arm independently.  The parameters are summarised in
Table~\ref{tab:params} and the radial profile shown in
Fig.~\ref{fig:r_prof}.

%%%%%%%%%%%%%%%
%%%%%%%%%%%%%%%

%%%%%%%%%%%%%%%
\begin{figure}
%work/galactic_b/broadbent/plane_mcmc/plots/model_best2.ps
%\includegraphics[width=\linewidth]{figs/model_best2.epsi} \caption{An example 
\includegraphics[width=\linewidth]{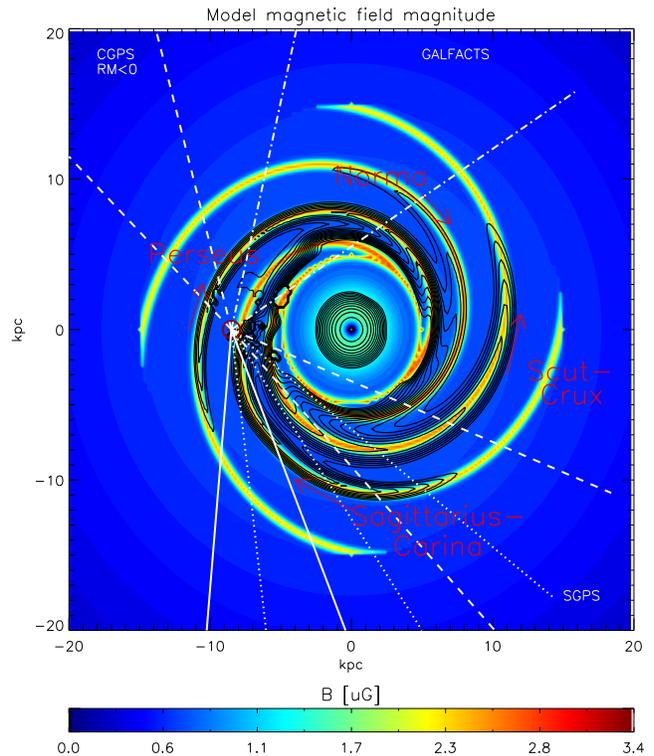} 
\caption{An example 
of a coherent spiral arm magnetic field model (no random component shown)
compared to NE2001 spiral model for the thermal electron density shown
as overlaid contours.  Interesting sight-lines are over-plotted
corresponding to the vertical lines in Fig.~\ref{fig:plane_dataonly}.
The peaks in the RMs correspond to looking roughly tangentially to a
spiral arm.  (Note that the contours do not make clear the
high-density ``molecular ring'' between roughly 2 and 5~kpc of the
Galactic centre.)  For comparison with Table~\ref{tab:params}, the
Perseus arm corresponds to amplitude $a_0$, Sag-Carina with $a_1$,
Scut-Crux $a_2$, Norma $a_3$, and the ring $a_4$.  (As discussed in
\S~\ref{sec:reg_prof}, the parameters used in this example are {\em not}
necessarily the defaults in that table.)}
\label{fig:regmodel}
\end{figure}
%%%%%%%%%%%%%%%

%%%%%%%%%%%%%%%
\subsection{\bf Random magnetic field}

The polarisation of synchrotron emission in the spiral arms of
external galaxies is generally low (a few percent according to
\citealt{beck:2009}), implying that a large fraction of the total 
magnetic field is due to an isotropic random component.

The random component we simulate begins with a Gaussian Random Field
(GRF) simulated in Fourier space ($k$ is the wavenumber, $1/\lambda$)
with a variance given by $|\mbold{k}|^{(\alpha-2)/2}$, where $\alpha$
is the spectral index of the power law describing the total magnetic
energy in one dimension, $P(k)\propto k^\alpha$, following
\citet{han:2004}.  For a Kolmogorov-type simulation that perhaps
matches the small-scale ($1/k\ltsim 1$pc) fluctuations, $\alpha=-5/3$,
while \citet{han:2004} find an index of $\alpha=-0.37$ at larger
($1/k\gtsim 1$kpc) scales.

This field in $k$-space is then Fourier transformed to real space and
amplified along the ridges of magnetic spiral arms by a parametrized
compression described in the next section.

%Note that the code we use (described below) includes a function to
%generate the GRF (and modify it as described in the next section), but
%it can also read in any otherwise-generated random field, \eg from a
%3-D MHD simulation.  

In this paper, we are simulating the entire galaxy, though only in two
dimensions.  In \S~\ref{sec:testing}, we describe how the resolution
of the random component affects the results.  In short, we find that
the analysis of the whole plane at low resolution is not strongly
sensitive to many parameters of the GRF.  We use the spectral index
for larger scales, $\alpha=-0.37$, a box size 40~kpc long on each
side, and 512 bins for a bin size of 80~pc.  This is very low
resolution and not much better than a single scale random component.
But as described in \S~\ref{sec:testing}, we find that increasing the
resolution does not change the results significantly.

For later papers, we will analyse small regions of the sky at high
resolution (\eg pixels of $0.5$~arcmin) and look in detail at the properties
of the turbulent component, but for this paper, those details are not
constrained.

%%%%%%%%%%%%%%%
\subsection{\bf Compression}\label{sec:compression}

As discussed above, in some external galaxies, the magnetic field
appears enhanced in the spiral arms as traced by the DIG, while in
others, there appear to be separate ``magnetic arms'', sometimes
between the DIG arms or showing a different (even varying) pitch
angle.  We therefore define a logarithmic spiral arm model for such
ridges parallel to the coherent field direction and allow the
orientation of the spiral to vary to fit the data.  In addition to the
spiral arms, we also include an annulus at 5~kpc from the Galactic
centre, roughly coinciding with the ``molecular ring'' feature in the
NE2001 model.

We define a four-arm model in this paper, and will test other models
in a later work.  Each arm is then described by a curve
\beq
r(\phi)=R_s \exp\left[(\phi_{0\rmn{n}}-\phi)/\beta\right]
\eeq
where $\beta\equiv 1/\tan(\theta_p)$, $\theta_p$ being the same pitch angle defined
above for the coherent field direction.  The angle $\phi_{0\rmn{n}}$ is
the reference angle for the $n^\rmn{th}$ spiral arm, always $\pi/2$
away from its neighbours in the case of four arms.

The arm ridges (and ring) are then defined by a modulation that peaks
along the ridge and reaches a minimum between the arms.  The
compression profile we use is based on \cite{broadbent:1990}, who
modelled a magnetic field component that is enhanced in the arm by a
compression factor $\rho_c$,
\beq\label{eq:rhoc}
%\rho_c = c(r) \exp(-(d/d_0)^2)+1
\rho_c = C_0 \exp(-(d/d_0)^2)+1
\eeq
where $d$ is the distance to the nearest of the four arm ridges as
measured along the line passing through the Galactic centre.  
%The
%amplitude of this modulation, $c(r)$ is constant in the inner galaxy
%and drops linearly to zero in the outer galaxy.  I.e.,
%\beq
%c(r)= \begin{cases}
%	C_0 & \mbox{if } r < R_3 \\
%	C_0 \left[1-(r-R_3)/(R_\rmn{max}^\rmn{arms}-R_3)\right] & \mbox{if } R_3 < r < R_\rmn{max}^\rmn{arms} \\
%	0 & \mbox{ if } r>R_\rmn{max}^\rmn{arms} \\
%\end{cases}
%\eeq

This implies a field that is unaltered between the arms but amplified
in the arms according to a Gaussian profile, when seen though a radial
cross-section, with a width of $d_0$.  This amplitude enhancement
profile applies to both the isotropic random component and the coherent
component.  This Gaussian modulation may not accurately represent what
may in reality be a shock front with a different profile, but it is
meant simply as a first order approximation to an enhancement parallel
to the shock plane.  (Though we refer to the ``shock plane'', note
that $\rho_c$ is not the density contrast across the shock, as often
used, but simply the total Gaussian density enhancement as compared to
the un-shocked inter-arm region.)  An example of the compressed
coherent field viewed from above is shown in Fig.~\ref{fig:regmodel}.

To simulate an ordered component, the random field is then also {\it
stretched} along the plane defined by the arm ridge.  This can be
thought of as an anisotropic random component, or as the addition of
an ordered component (in addition to the coherent and isotropic random
components) whose orientation is always along the spiral arm but whose
specific direction changes stochastically.  Rather than using the
numerical approximation given in Broadbent et al., we apply this
compression explicitly to our simulated random component.  Namely, we
use
\beq\label{eq:ford}
%{\bf B}_\rmn{irreg}=\rho_c\left({\bf B}_\rmn{GRF} +  f_\rmn{ord} (\rho_c-1){\bf B}_\rmn{GRF}^\rmn{proj}\right) \equiv {\bf B}_\rmn{iso} + {\bf B}_\rmn{ord}
{\bf B}_\rmn{irreg}=\rho_c{\bf B}_\rmn{GRF} +  f_\rmn{ord} (\rho_c-1){\bf B}_\rmn{GRF}^\rmn{proj} \equiv {\bf B}_\rmn{iso} + {\bf B}_\rmn{ord}
\eeq
where ${\bf B_\rmn{GRF}^\rmn{proj}}$ is the component of the random
field projected onto the shock plane,
\beq
{\bf B}_\rmn{GRF}^\rmn{proj} = {\bf B}_\rmn{coh} \frac{ {\bf B}_\rmn{coh}\cdot{\bf B}_\rmn{GRF}}{|{\bf B}_\rmn{coh}|^2} ,
\eeq
and $f_\rmn{ord}$ sets the ratio of the ordered field to the amplified
but isotropic random component in the arms.  (Note that $f_\rmn{ord}$
is not the ratio itself, since that changes from the arm to inter-arm
regions.) The first term in Eq.~\ref{eq:ford} is then the isotropic
(but no longer homogeneous) random component while the second term is
the ordered component.  The shock plane is assumed to be perpendicular
to the Galactic plane and parallel to the spiral arm traced by the
coherent component.  

The form of this parametrization means that in the inter-arm regions,
the ordered component goes to zero, while the isotropic random
component is not amplified but non-zero.  The combination of
$B_\rmn{RMS}$ and $C_0$ determines the amplification in the arm
relative to the inter-arm region.  In particular:
\beq
\left<B_\rmn{iso}^2\right>^{1/2}= \begin{cases}
	\left<B_\rmn{GRF}^2\right>^{1/2}\equiv B_\rmn{RMS}, & \mbox{as } \rho_c \rightarrow 1 \mbox{ (interarm)} \\
	(C_0+1) B_\rmn{RMS}, & \mbox{as } \rho_c \rightarrow C_0+1 \mbox{ (ridge)}. \\
\end{cases}\label{eq:peaks}
\eeq
In other words, in the inter-arm region, $d$ is large, $\rho_c$ goes
to $1$ (see Eq.~\ref{eq:rhoc}).  On the ridge, $d$ goes to zero,
$\rho_c$ reaches its maximum at $C_0+1$, and the random component is
amplified by $C_0+1$.  In between, the amplification is falling off
with a Gaussian profile of width $d_0$.  The profile of the ordered
component, proportional to $\rho_c-1$, is the same Gaussian but
without the offset, \ie it goes to zero between the arms:
\begin{eqnarray}
\lefteqn{\left<B_\rmn{ord}^2\right>^{1/2} =}  \notag \\
& \begin{cases}
	0, &  \mbox{as } \rho_c \rightarrow 1 \mbox{ (interarm)} \notag \\
%	f_\rmn{ord}(C_0+1)C_0\left<(B_\rmn{GRF}^{proj})^2\right>^{1/2} & \\
%	\mbox{   } =f_\rmn{ord}C_0 \sqrt{\frac{2}{3}}\left<B_\rmn{iso}^2\right>^{1/2}, & \mbox{as } \rho_c \rightarrow C_0+1  \mbox{ (ridge)} \\
	f_\rmn{ord}C_0\left<(B_\rmn{GRF}^\rmn{proj})^2\right>^{1/2} & \mbox{as } \rho_c \rightarrow C_0+1  \mbox{ (ridge)} \\
\end{cases} \\
\end{eqnarray}
Note that this model approximates the case where the ordered component
is due to the compression wave in the arm, but it does not represent
the possibility that the ordered component might arise simply due to
differential rotation.

The ratio of ordered to random is then zero in the inter-arm regions;
along the ridge, the ratio is 
\beq\label{eq:ratio}
\frac{ \left<B_\rmn{ord}^2\right>^{1/2} } { \left<B_\rmn{iso}^2\right>^{1/2} } = f_\rmn{ord}\frac{C_0}{C_0+1}\sqrt{\frac{2}{3}}
\eeq
because
$\left<(B_\rmn{GRF}^\rmn{proj})^2\right>=2/3\left<B_\rmn{GRF}^2\right>$.

%$f_\rmn{ord}C_0\sqrt{2/3}$.
%

%
%Therefore, we have
%\beq
%{\bf B'}=\rho_c\left[{\bf B} + (\rho_c-1)\frac{|{\bf B}_{ran}}|}{|{\bf B}_{coh}}|}\cos\alpha{\bf B}_{coh}}\right]
%\eeq
%where $\alpha$ is the angle between the coherent and random components.

%When looking down a spiral arm, ${\bf B}_{coh}}={\bf B_{coh}^\parallel}$.  
%So we end up with: 
%%
%$${\bf B'_{ran}}=\rho_c{\bf B_{ran}^\perp} +\rho_c^2{\bf B_{ran}^\parallel}$$
%%

When looking down a spiral arm, the ordered component does not
contribute to the synchrotron emission.  The impact of this change on
total intensity is then seen in the {\em relative} amount of emission
seen looking down an arm relative to elsewhere, \ie in the shape the
emission steps seen in the longitude profiles in
Fig.~\ref{fig:plane_dataonly}.  Since the isotropic component does not
contribute to polarisation, the ordered component, along with the
coherent field, also determines the polarised emission profile.

The same effect on the synchrotron can be obtained with an isotropic
compression of the random component combined with a stronger coherent
component.  Amplifying the random component along the arms essentially
adds an ordered component, which as described above, is
indistinguishable from a coherent component in synchrotron.  The two
cases can be distinguished, however, by looking at the rotation
measures as well, where a too-large coherent component will overpredict
RM.

%%%%%%%%%%%%%%%
\subsection{\bf Thermal electrons}\label{sec:ne}

For the thermal electrons, we consider three possibilities:
\begin{itemize}
\item a physically unrealistic constant density;
\item the NE2001 model;  or
\item a parametrized spiral arm structure analogous to the coherent 
magnetic field component.
\end{itemize}

The thermal electron density model that has been most widely used is
that of \citet{cordes:2002}, known as NE2001.  Based on observations
of pulsars at known distances, this model relates the thermal electron
density, $n_e$, to the dispersion measure (DM) of a give source and
its distance, $D$:
\beq\label{dm}
DM\equiv\int_0^D n_e \rmn{d}l
\eeq
Though this model was intended for determining pulsar distances, it is
also used for modelling the magnetic field through RM, as it is the
best existing predictor of the thermal electron density.
%, though that may be largely a testament to the need
%for a model rather than the accuracy of NE2001 in detail.  The
%authors intended it as a tool to predict pulsar distances, but it is
%only first order as a predictor of, \eg free-free emission from
%thermal electrons.

\begin{figure}
\includegraphics[width=\linewidth]{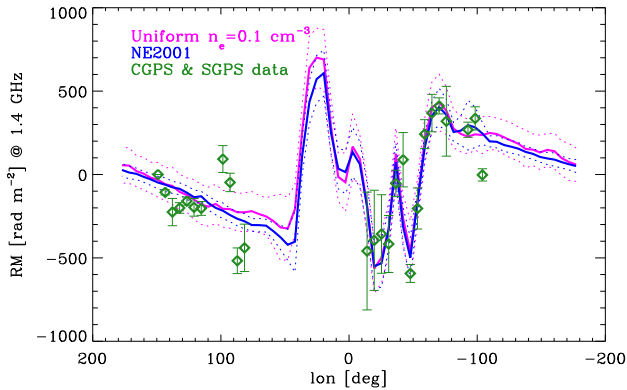}
%  originally from
%  work/galactic_b/plane_mcmc/try_data2
% IDL> plot_ne2001,['0','0'],roots=['../plotting/result_6para_constant/out/','../plotting/result_6para_ne2001/out/'],psfile='../plots/compare_ne2001.eps'
%
\caption{A comparison of the rotation measures from two models fitted to the 
CGPS and SGPS data, in the one case with a constant thermal electron
density of $n_\rmn{e}=0.1$ and in the other case with the NE2001
prediction.  The best-fit parameters are given in Table
\ref{tab:results}.  (The dotted lines indicate the galactic 
variance due to the random magnetic field component.)}
\label{fig:compare_ne2001}
\end{figure}

The constant model is useful as a comparison to determine whether or
not we can distinguish either of the others from the null hypothesis.
We have compared the results of using a constant density to those
using the NE2001 model and find that the resulting data profiles are
largely indistinguishable.  The models and RM data are shown in
Fig.~\ref{fig:compare_ne2001}, with the corresponding parameters that
were fitted listed in Table~\ref{tab:results}.  (The synchrotron
profiles, not shown, are effectively indistinguishable.)  The curves
are essentially the same, and though the individual arm amplitudes
change, the effect on the synchrotron emission would be very difficult
to distinguish.

In this work, we use the NE2001 thermal electron density model unless
otherwise specified.  Though it is not likely to be correct in detail,
it is at least based on observations and provides the best current
estimate.

The possibility of defining our own model introduces quite a few more
parameters to the problem, and those are unlikely to be constrained
with the data we are using.  In the future, however, we can include
dispersion measures that depend only on the thermal electron density
(and are calculated in the integration code we use) and additional
data about the distribution from H$\alpha$ emission.  The tools we use
can be extended to include galactic pulsars, comparing that additional
RM and DM data to better map out the thermal electron distribution
near the Sun.  As each pulsar is at a different distance, this
complicates the LOS integration and the likelihood exploration, but in
future work, we can explore this idea further.

%%%%%%%%%%%%%%%
\subsection{\bf Cosmic-ray electrons}\label{sec:cres}

The spatial and spectral distribution of cosmic rays in the Galaxy is
thought to be fairly smooth until energies above ~100~GeV (see,
\eg \citealt{strong:2004} and \citealt{strong:2007}).  But it is not 
clear {\it how} smooth the distribution is, and local measurements are
obviously affected by local sources, so it is difficult to determine
the average distribution over the galaxy.  

For the purposes of constraining the structure of the magnetic field,
however, we start simple.  We adopt an exponential disc model such as
that used by \citet{page:2007} and
\citet{sun:2008} (motivated by the work of \citealt{drimmel:2001}), namely
\beq\label{eq:cre}
J_\rmn{CRE}(r,z) = J_\rmn{CRE,\oplus}
\exp(-(r-R_\oplus)/h_r)
\eeq
where $R_\oplus$ is the Galactocentric radius of the Solar system.
(In the current analysis on the plane only, we neglect any modulation
with height.)  The normalisation is uncertain, but as a first
estimate, we read off the value at 10~GeV from Fig.~4 of
\citet{strong:2007} of roughly $E^3 J(E=10\,\rmn{GeV})
\approx 250\,\rmn{GeV}^2\rmn{m}^{-2} \rmn{s}^{-1} \rmn{sr}^{-1}$.  We then 
use the value of
$J_\rmn{CRE,\oplus}=0.25\,\left(\rmn{GeV\,m^2\,s\,sr}\right)^{-1}$ to
set the cosmic ray normalisation.  
\footnote{
This is sometimes expressed as a spatial number density normalisation.
The spatial number density (assuming relativistic electrons with
$v\sim c$) is then $N(E)=J(E)\times 4\pi
\times 1/c $.  Often, what is given is a normalisation $C(r,z)$ such that 
$N(\gamma)d\gamma\equiv C(r,z) \gamma^{-p}d\gamma$ where $\gamma$ is
the Lorentz factor ($E\equiv\gamma m_\rmn{e}c^2$).  Therefore,
$C_\oplus=N(\gamma)\gamma^p=J\times \frac{4\pi}{c}\times
m_\rmn{e}c^2\times \left(\frac{10\, \rmn{GeV}}{m_\rmn{e}c^2}\right)^3=
4\times 10^{-5}\, \rmn{cm}^{-3}$, assuming $p=3$.}
For comparison, \citet{sun:2008}
use a value of
$J_\rmn{CRE,\oplus}=0.4\,\left(\rmn{GeV\,m^2\,s\,sr}\right)^{-1}$.

For this work, we also adopt a simple power law energy distribution
with the index $p=3$.

We should emphasize that this model is overly simplistic in assuming a
smooth spatial distribution and a simple power law spectrum that is
the same throughout the Galaxy.  Neither of these assumptions is
likely to be true in reality.  What we are constraining is then
the spatial distribution of the product $n_\rmn{CRE}B^2$.  The
degeneracy between spatial variations in the two independently is only
broken for the coherent field using the rotation measure data.  In a
later work, we will see how much the two can be separated by adding
thermal emission from dust to the analysis.

\begin{table*}
%\begin{center}
\begin{center}
\begin{tabular}{ | l | l | l | p{6cm} |}	
%\begin{tabular}{ | l | l | l | p{5cm} |}

%--------------------------------------
\hline
Parameter & Default & Equation & Description \\
\hline
\multicolumn{4}{c}{{\bf Coherent magnetic field:}  $B_{coh}(r,\phi)=B_0(r)a_n\rho_c(d)$} \\
\hline
$B_0$ & 1 $\mu$G & &  Global amplitude normalisation\\
$R_s$ & 7.1 kpc & see $p$ & Scale radius of spiral.\\
$\theta_p$ & -11.5 & $\beta\equiv 1/\tan(\theta_p)$ and &  $\theta_p$ is the pitch angle of spiral \\
 & &  $r(\phi)=R_s \exp\left[(\phi_0-\phi)/\beta\right]$ & ($r(\phi)$ gives arm radius at given azimuth)\\
$\phi_0$ &  & See pitch & Angle representing rotation of spiral around axis through Galactic poles \\
$R_\rmn{mol}$ & 5. kpc & & Radius of ``molecular ring'' \\
$R_\rmn{max}$ & 20 kpc & & Maximum radius, beyond which $|B|=0$\\
%$R_\rmn{max}^\rmn{arms}$ & 15 kpc & & Maximum radius, beyond which $c(r)=0$ (see $C_0$).\\
$R_\rmn{max}^\rmn{arms}$ & 15 kpc & See $C_0$ & Maximum radius, beyond which $C_0=0$ (see $C_0$).\\
$N_\rmn{arms}$ & 4 & & Number of spiral arms \\
$a_n$ &  & $B_{coh}(r,\phi)=B_0(r)a_n\rho_c(d(r,\phi))$ & Amplitude modulation and direction for a given arm, $n$.\\
%$d_0$ & 0.3 kpc & $\rho_c(d)= c(r) \exp(-(d/d_0)^2)+1$ & Defines width of arm for density enhancement, $\rho_c$;  $d$ is the distance to the nearest arm in kpc \\
$d_0$ & 0.3 kpc & $\rho_c(d)= C_0 \exp(-(d/d_0)^2)+1$ & Defines width of arm for density enhancement, $\rho_c$;  $d$ is the distance to the nearest arm in kpc \\
%$R_0$ & $13.2$ kpc & $B_0(r)=B_0(1-\exp(-r^2/r_2^2))(\exp(-r^2/R_0^2)+\exp(-r^4/R_1^4))$ & outer radial profile parameter \\
$R_0$ & 20 kpc & $B_0(r)=B_0(1-\exp(-r^2/R_2^2))(\exp(-r^2/R_0^2)+\exp(-r^4/R_1^4))$ & Outer radial profile parameter.  See Fig.~\ref{fig:r_prof}. \\
%$R_1$ & $2.25$ kpc & see $R_0$ & inner radial profile parameter \\
$R_1$ & 3 kpc & see $R_0$ & Inner radial profile parameter \\
$R_2$ & 0.5 kpc & see $R_0$ & \\
$C_0$ & 1 & See $d_0$.  $C_0 = 0 \mbox{ if } r > R_\rmn{max}^\rmn{arms}$ &  Peak density contrast.  Sometimes $C_0\propto a_n$;  see \S~\ref{sec:scaling}.\\ 
%$C_0$ & 1 & $c(r)=C_0 \mbox{ if } r < R_3$ & \\
%	& & \hspace{1cm}$C_0 \left[1-(r-R_3)/(R_\rmn{max}^\rmn{arms}-R_3)\right] \mbox{ if } R_3 < r < R_\rmn{max}^\rmn{arms}$ \\  
%        & & \hspace{1cm}$0 \mbox{ if } r > R_\rmn{max}^\rmn{arms}$ & Peak density contrast.  Sometimes $C_0\propto a_n$;  see \S~\ref{sec:scaling}.\\ 
%$R_3$ & 15 kpc & see $C_0$ & \\
%$\phi_b$ & ? & ? & Orientation of bar \\
%$e$ & ? & ? & Axis ratio of bar \\

%--------------------------------------
\hline\multicolumn{4}{c}{\bf Random and ordered magnetic field}  \\
\hline
$\alpha$ & -0.37 & $P_B(k)\equiv \left<B_\rmn{ran}(k)^2\right>\propto k^\alpha$ & Power law spectral index of initial GRF;  default from \citet{han:2004} in 1D, Kolmogorov value $-5/3$. \\
$D_\rmn{co}$ & 1 kpc & $B_\rmn{ran}(k)=0$ for $k<1/D_\rmn{co}$ & Cutoff maximum of GRF fluctuations (minimum determined by resolution) \\
$B_\rmn{rms}$ &  & $B_\rmn{rms}\equiv \left<B_\rmn{ran}^2({\bf x})\right>^{1/2}$ & Total RMS amplitude of GRF fluctuations \\
$f_\rmn{ord}$ &  & ${\bf B'}_\rmn{ran}=\rho_c{\bf B}_\rmn{ran} +  f_\rmn{ord} (\rho_c-1){\bf B}_\rmn{proj} \equiv {\bf B}_\rmn{iso} + {\bf B}_\rmn{ord}$  & Relates ordered to isotropic random component \\

%--------------------------------------
\hline
\multicolumn{4}{c}{\bf Thermal electrons}\\
\hline
$n_\rmn{e0}$ & - & $n_\rmn{e}(r,\phi)=n_\rmn{e0}$ & Alternative constant density test model.  By default, we use the NE2001 model of \citet{cordes:2002}, which peaks in the arms around $0.1\,\rmn{cm}^{-3}$.\\

%--------------------------------------
\hline
\multicolumn{4}{c}{\bf Cosmic-ray electrons} \\
\hline
$p$ & 3 & & Electron power spectrum power law index.  See \S~\ref{sec:synch} and \S~\ref{sec:cres}. \\
%$p$ & 3 & $n_\rmn{cre}(r,z)=n_\rmn{cre,0} \exp{\left(-r/h_r\right)} \rmn{sech}^2\left(z/h_d\right)$ & Model used in \citet{page:2007} \\
$h_r$ & 15 kpc & $J_\rmn{CRE}(r) = J_\rmn{CRE,\oplus} \exp(-(r-R_\oplus)/h_r)$ & Scale radius of CREs. \\
%$h_z$ & 1 & & \\
%$C_\rmn{CRE,\oplus}$ &   & $C_\rmn{CRE}(r=R_\oplus,z=0)=4\times 10^{-5} \rmn{cm}^{-3}$ & See \S~\ref{sec:cres}. \\
$J_\rmn{CRE,\oplus}$ & $\frac{0.25}{\rmn{GeV}\,\rmn{m}^2\,\rmn{s\,sr}}$ & 
%$C_\rmn{CRE,\oplus}=J(10\,\rmn{Gev})\frac{4\pi}{c} m_\rmn{e}c^2\left(\frac{10 \rmn{GeV}}{m_\rmn{e}c^2}\right)^3 =4\times 10^{-5} \rmn{cm}^{-3}$ 
& See \S~\ref{sec:cres}. \\

%--------------------------------------
\hline
\end{tabular}
\end{center}
\caption{Table of modelling parameters as described in \S~\ref{sec:models}.}\label{tab:params}
\end{table*}

%%%%%%%%%%%%%%%%%%%%%%%%%%%%%%%%%%%%%%%%%%%%%%%%%%%%%%%%
\section{Model Selection Method}\label{sec:method}
%%%%%%%%%%%%%%%%%%%%%%%%%%%%%%%%%%%%%%%%%%%%%%%%%%%%%%%%

The problem is then to parametrize the Galactic magnetic field (as
well as the the thermal and cosmic ray electrons if possible) and find
the model parameters that best fit the data.  There are, then, a large
number of parameters, and a brute-force approach like a simple grid
search would require prohibitive amounts of computing time.  Instead,
we use the far more efficient and flexible method of a Monte Carlo
Markov Chain (MCMC) analysis.

\subsection{Simulation with {\sc hammurabi}}

The likelihood function we use computes a simple $\chi^2$ from a
comparison of the model to the data along the Galactic plane as shown
in Fig.~\ref{fig:plane_dataonly}. For the given values of the input
parameters, we generate the observed emission and RM at each pixel by
performing a line-of-sight integration through a simulated galaxy
using the {\sc hammurabi} code of \citet{waelkens:2009}.

The {\sc hammurabi} code is designed to compute observables such as
synchrotron emission and dust emission in full Stokes parameters while
taking into account Faraday rotation and depolarisation effects.  It
performs a line-of-sight integration through a galaxy simulation, but
uniquely, it can refine the integration resolution as the distance
increases in order to maintain a roughly constant physical cell size.
This is crucial for simulating effects such as beam depolarisation.
We do not take full advantage of this code in this current work, but
we will later use it to investigate more thoroughly the properties of
the turbulent ISM and its effect on the observables.

For this analysis, we simply use one observing ``shell'' (\ie no grid
refinement with distance) and a simple 2D simulation in the plane of
the Galaxy only.  We compute the observables at each of 512 pixels
along the plane that correspond to the centres of the HEALPix pixels
on the plane in a map at $N_\rmn{side}=128$ map.  We then smooth the
result in the same way we smoothed the data as described in
\S~\ref{sec:synch} and bin them into 64 pixels.

\subsection{MCMC with {\sc cosmomc}}

For the MCMC sampling, we use the tools already developed by
\citet{lewis:2002} and publicly available as the package {\sc
cosmomc}.  It consists of a Fortran sampling routine, into which one
can insert a parametrized likelihood function, and an analysis tool,
{\it getdist}, to read the resulting Markov chains and determine the
parameters' mean and maximum likelihood values and correlation
matrices.  We use the {\sc cosmomc} sampler but have written our own
analysis tool in order to add further functionality described below.
The sampler has the ability to continuously update the proposal
density based on the likelihood space so far sampled, which makes for
a more efficient mapping out of the likelihood.

This method is complicated in this case by the fact that we do not
compute the model, \ie the simulated emission and rotation measures,
analytically.  Rather, as described above, we simulate a Gaussian
random field for the random component and apply the compression
algorithm.  The model we wish to compare to the data is then the
expectation value of the emission profiles resulting from such
simulations, and the corresponding variance among realisations
determines the error.  These are difficult to compute analytically, so
we instead use a brute-force approach of creating a set of $N=10$ 
realisations and taking the mean.  This must be done for each
likelihood evaluation.  (We have tried several different numbers of
realisations at each sample and found ten to be the most efficient in
terms of the trade-off between sample variance and run time.)

\begin{figure}
\includegraphics[width=0.8\linewidth]{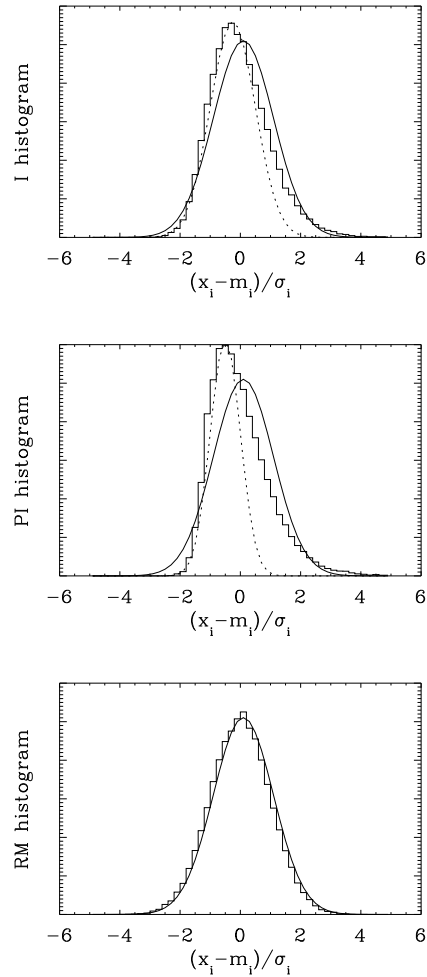}
\caption{ 
Distribution of data values compared to the mean.  For 1000
simulations and 64 pixels along the plane, the histogram shows the
difference between the data for a given simulation and at a given
pixel, $x_i$, and the mean at that pixel over all simulations, $m_i$,
normalised by the variance, $\sigma_i$.  Clearly, for total and
polarised intensity, there is a tail on the positive side that biases
the mean.  The solid line shows a Gaussian of the equivalent mean and
width, which is essentially what using a $\chi^2$ assumes, while the
dotted line shows a Gaussian that approximately fits the left half of
the histogram.  }
\label{fig:pixel_dists}
\end{figure}

For simplicity, we assume a Gaussian distribution of the data points,
$x_i$, about the mean model value, $\mu_i$, with a variance of
$\sigma_i$ so that the likelihood is:
\beq
-\ln{\left(\mathcal{L}\right)}=\sum_i \ln{ \left(\sqrt{2\pi}\sigma_i\right) } + \left(x_i - \mu_i\right)^2/(2 \sigma_i^2 )
\label{eq:like}
\eeq
(where the summation is over the pixels along the plane).  The
distributions are not exactly Gaussian, as can be seen in
Fig.~\ref{fig:pixel_dists}, but the assumption appears to work
reasonably well nonetheless.  The differences come from the relation
between the Gaussian random field and the observables, which in the
case of the synchrotron emission, is non-linear.  The RMs of
extragalactic sources start with an average value through the galaxy,
dependent only on the coherent component of the magnetic field, that
is then perturbed by a one-dimensional random walk due to the random
magnetic field component projected along the LOS (assuming the
small-scale fluctuations are uncorrelated with the thermal electrons).
The random component then adds a variance proportional to the number
of steps.  (Because the random component has structure on many scale
lengths, it is essentially a superposition of random walks with
different sized steps.)  For polarised intensity, the polarisation
vector behaves like a two-dimensional random walk, and the observable
is only the length of the result.  With only a GRF, the expectation
value of either Stokes Q or U would be zero, but the expectation of
$P=\sqrt{Q^2+U^2}$ is non-zero and adds to the coherent component.
For total intensity, the emission of all components simply adds.

Note that the variance at each longitude depends both on the position
and on the model parameters and is determined simply from the
variance of the simulated realisations.  There is an additional
uncertainty in both the model mean, $\mu_i$, and its variance,
$\sigma_i$, due to the limited number of realisations, so in the
determination of $\chi^2$, we actually use
\beq
\hat{\sigma}_i^2=(1+1/N)\sigma_i^2
\eeq
in the likelihood evaluation of Eq.~\ref{eq:like}.

%%%%%%%%%%%%%%%%%%%%%%%%%%%%%%%5
%work/galactic_b/plane_mcmc/test_lowres/test_distributions/scripts/read_sims.pro
%%%%%%%%%%%%%%%%%%%%%%%%%%%%%%%

When analysing the chains, there is essentially an additional
``noise'' term added to the likelihood space, since two evaluations at
the same point in parameter space will return slightly different
likelihoods.  This means that we need more samples to overcome the
sample variance and to map accurately the shape of the likelihood
space.  It also means that we need to give a ``temperature'' parameter
to the sampler.  The latter is necessary to prevent the algorithm from
falling into an unusually low $\chi^2$ hole and never getting out
again.  So rather than sampling from $P\propto \exp{-\chi^2/2}$, we
sample from $P\propto \exp{-\chi^2/2T}$, where $T$ is not a physical
temperature but simply a way to adjust the sampler. This temperature is
simply tuned to ensure that the acceptance rate of the sampler (\ie
the fraction of proposed samples that are accepted) remains roughly a
third.  (With a total of 140 data points and corresponding $\chi^2$
values, we find a temperature of 10 usually sufficient.)  This
temperature correction is then reversed to analyse the results,
correcting both the likelihoods as well as the weights (the number of
steps in the chain where it remained at a given sample before it
accepted the next).

To simplify the analysis of the chains, we need to remove this noise,
and we do this simply by smoothing the likelihood space (as, for
example, done in \citealt{dick:2006}).  For each sample, we replace
the likelihood with the mean of the nearby likelihoods.  We must also
smooth the weights, which are by construction meant to be proportional
to the likelihood.  The result of the smoothing is a distribution of
likelihood values approximating that seen in an essentially noiseless
situation with otherwise identical parameters.
%(We approximate this by fixing the random seed for the GRF generation,
%so that the ``noise'' is identical for every sample and the likelihood
%space remains relatively smooth, though the result is biased.)
We smooth by binning the samples in likelihood space with a binsize of
1/15 times the range (after burn-in) for each parameter.  

The number of samples required for the chains to converge (using {\sc
cosmomc}'s parameter {\tt MPI\_Limit\_Converge\_Err = 0.3}) is between
a few thousand for a 2D fit and a few hundred thousand for a 6D fit.

\begin{figure*}\begin{tabular}{cc}
%IDL> my2dh,'../../test_lowres/test_RMonly/chains/test1_',nchains=8,doparams=[4,5,6,7],savefile='../../test_lowres/scripts/my2dh_RMonly_idl.dat','../plots/my2dh_RMonly_sim_4params.eps',correct=[0.27,-0.54,0.22,-0.62],$
%IDL> charsize=2,plot_labels=['(a)','','(b)','(c)','(d)',''],doplots=[0,2,3,4]
\includegraphics[width=0.5\linewidth]{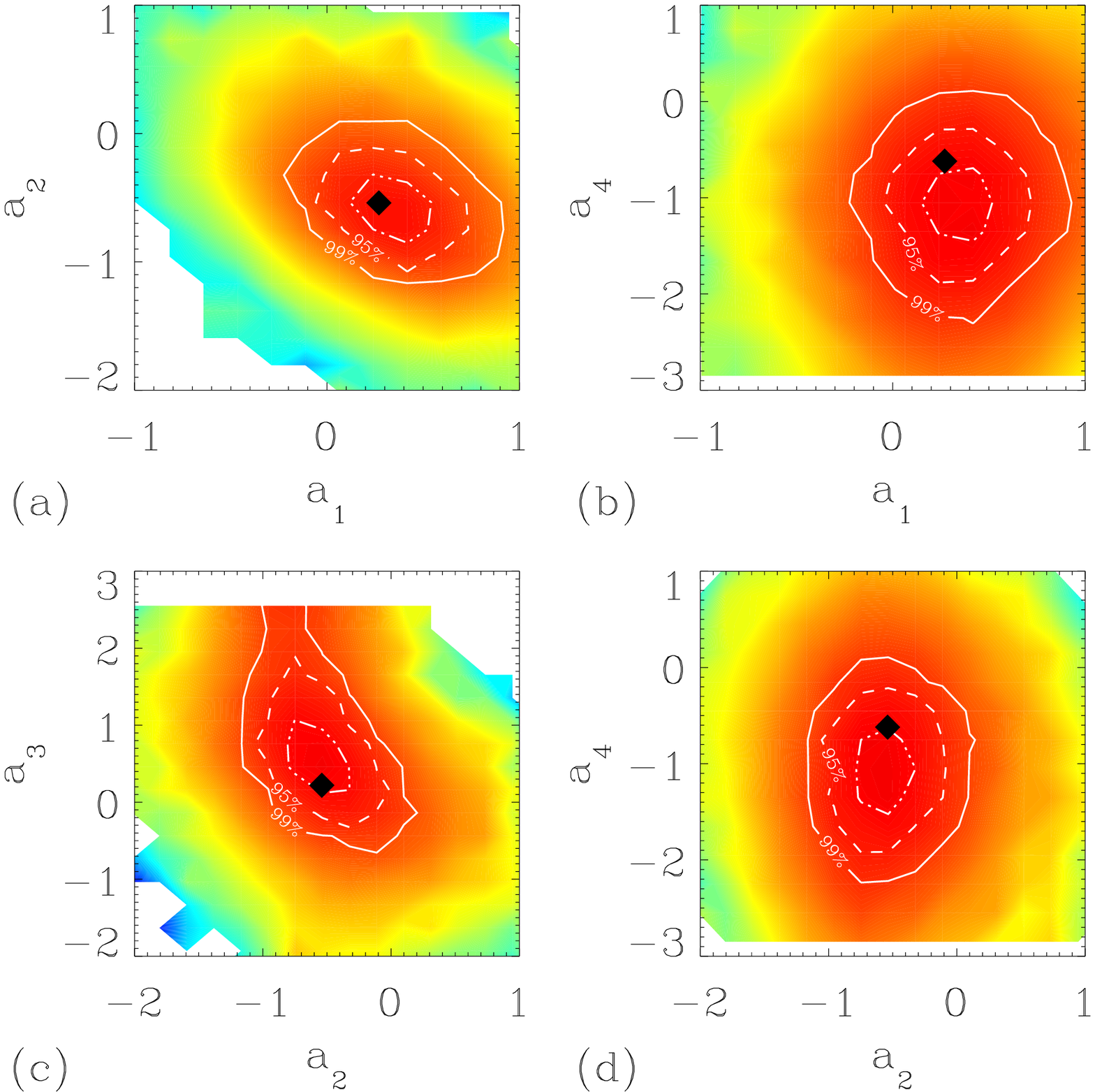} & 
% Originally from
% work/galactic_b/plane_mcmc/try_data2/plots/my2dh_sim4_para.eps
%
% IDL> my2dh,'../../test_lowres/sim_set/sim_4/chains/test1_',nchains=4,doparams=[2,8],correct=[0.4,1.5],xrange=[0,1.5],'../plots/my2dh_sim1_para.eps',plot_labels=['(e)']
%
\includegraphics[width=0.4\linewidth]{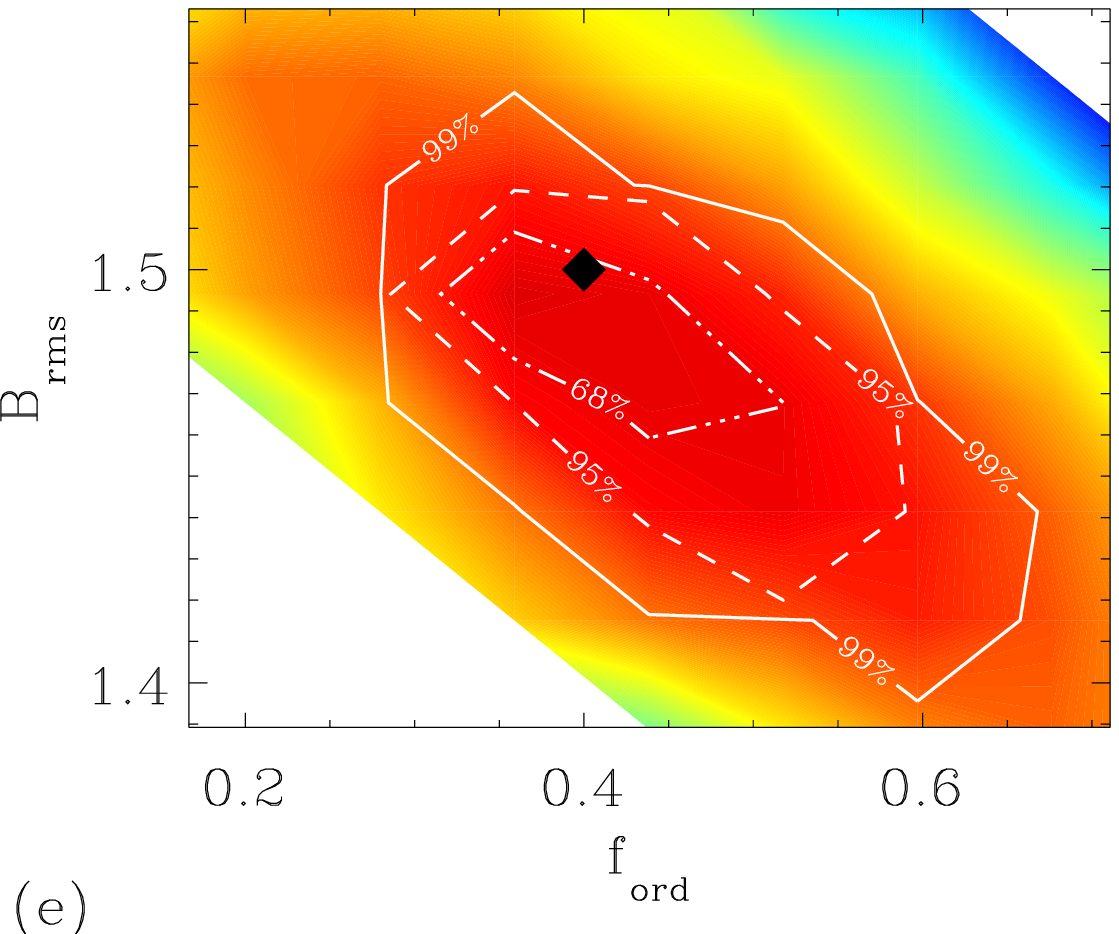} \\
\end{tabular}
\caption{Example of MCMC results.  The filled contours represent the mean 
likelihood showing the sampled region, while the white contour lines
indicate the 68, 95, and 99.7 per cent confidence regions.  The filled
black diamonds give the correct input values.  The examples on the {\it
left, (a) through (d),} show the results for several of the six
parameters describing the coherent magnetic field found fitting only
to the RM data.  On the {\it right (e)} are the results for the two
parameters describing the amount of isotropic random and ordered
components found using all the data.}
\label{fig:mcmc_example}
\end{figure*}

% work/galactic_b/plane_mcmc/test_case_coma
In \S\ref{sec:testing}, we describe how we tested with simulated
inputs to verify that the MCMC search returns the input parameters and
reasonable uncertainties.  An example is shown in
Fig.~\ref{fig:mcmc_example}.

\subsection{Parameters and datasets}

One could, in theory, throw all data and all parameters into the MCMC
analysis at once, but it would be surprising if such an approach
yielded useful information.  Degeneracies among parameters make
samplers highly inefficient, and when starting values are very far
from the peak likelihood, the burn-in time to random-walk over to the
peak can be very long.  Instead, we make educated guesses (informed by
trial and error) about what combinations of parameters and data will
give good constraints and test them using simulations.  For the
parameters we fix for the course of the analysis, we choose by hand
values that roughly match the data, since these are not our primary
interest.

For example, as described above, the RMs are the best dataset to
constrain the coherent magnetic field component.  Particularly, the
angle $\phi_0$ describing the azimuthal orientation of the spiral may
have a strong effect on the position of the RM sign reversals, while
the $a_n$ describe the relative magnitudes of the RM features.  While
these parameters do have an effect on the profiles of the synchrotron
emission, given the galactic variance, this may not be detectable.
So to determine these parameters, then, we use only the RM data.

The synchrotron emission is particularly interesting for studying the
different components of the random field.  As described above, we can
use the RMs to constrain the coherent field and then use the
synchrotron to constrain the parameters such as $B_\rmn{rms}$ and
$f_\rmn{ord}$, keeping $\phi_0$ and $a_n$ fixed to their best-fit
values from the RM-only analysis.

Table \ref{tab:params} lists all of the parameters that go into the
modelling, and clearly studying all of them would be a huge
undertaking.  For this first paper, we focus on the most immediately
interesting issues of the magnetic field reversals and the relative
contribution of the coherent, ordered, and random components.

%%%%%%%%%%%%%%%%%%%%%%%%%%%%%%%%%%%%%%%%%%%%%%%%%%%%%%%%
\section{Testing}\label{sec:testing}
%%%%%%%%%%%%%%%%%%%%%%%%%%%%%%%%%%%%%%%%%%%%%%%%%%%%%%%%

%%%%%%%%%%%%%%%%%%%%%%
\subsection{Basic Parameters}

Figure~\ref{fig:mcmc_example} shows an example of how the MCMC method
works with a simulated dataset.  The results on the left use only RM
data covering roughly half of the plane.   Among the six
parameters simultaneously fit for this test (all $a_n$ and $\phi_0$),
all return a mean, $\mu_i$, that is within $2\sigma$ of the input
value, $\mu_0$, where $\sigma$ is the variance of the samples.

On the right of Fig.\ref{fig:mcmc_example} is the result of the second
step of fitting.  We fix the amplitudes at the values found using the
RM data only and then fit to all data the two parameters controlling
the ratio of ordered to random magnetic fields in the arms.  The
returned values lie roughly within 1~$\sigma$ of the correct values.

% work/galactic_b/plane_mcmc/test_lowres/sim_set/scripts/idl_journal.pro.bak4
%
%IDL> print,mean([ 0.31,0.54,0.41,0.58,0.44,0.45,0.31,0.48,0.49,0.27]),sqrt(variance([ 0.31,0.54,0.41,0.58,0.44,0.45,0.31,0.48,0.49,0.27])/10.)
%     0.428000    0.0326531
% (correct value 0.4 +- 0.06 or so)
%
%IDL> print,mean([1.53,1.49,1.53,1.43,1.48,1.49,1.52,1.50,1.47,1.53]),sqrt(variance([1.53,1.49,1.53,1.43,1.48,1.49,1.52,1.50,1.47,1.53])/10.)
%      1.49700    0.0102252
% (correct value 1.5 +- 0.02 or so)
We have run a set of ten such simulations to verify that the estimates
of $f_\rmn{ord}$ and $B_\rmn{RMS}$ are correct and unbiased (\ie
$|\left<\mu_i\right> - \mu_0| < \sigma/\sqrt{10}$).

%%%%%%%%%%%%%%%%%%%%%%
\subsection{Resolution}\label{sec:reso}

\begin{figure}
\includegraphics[width=\linewidth]{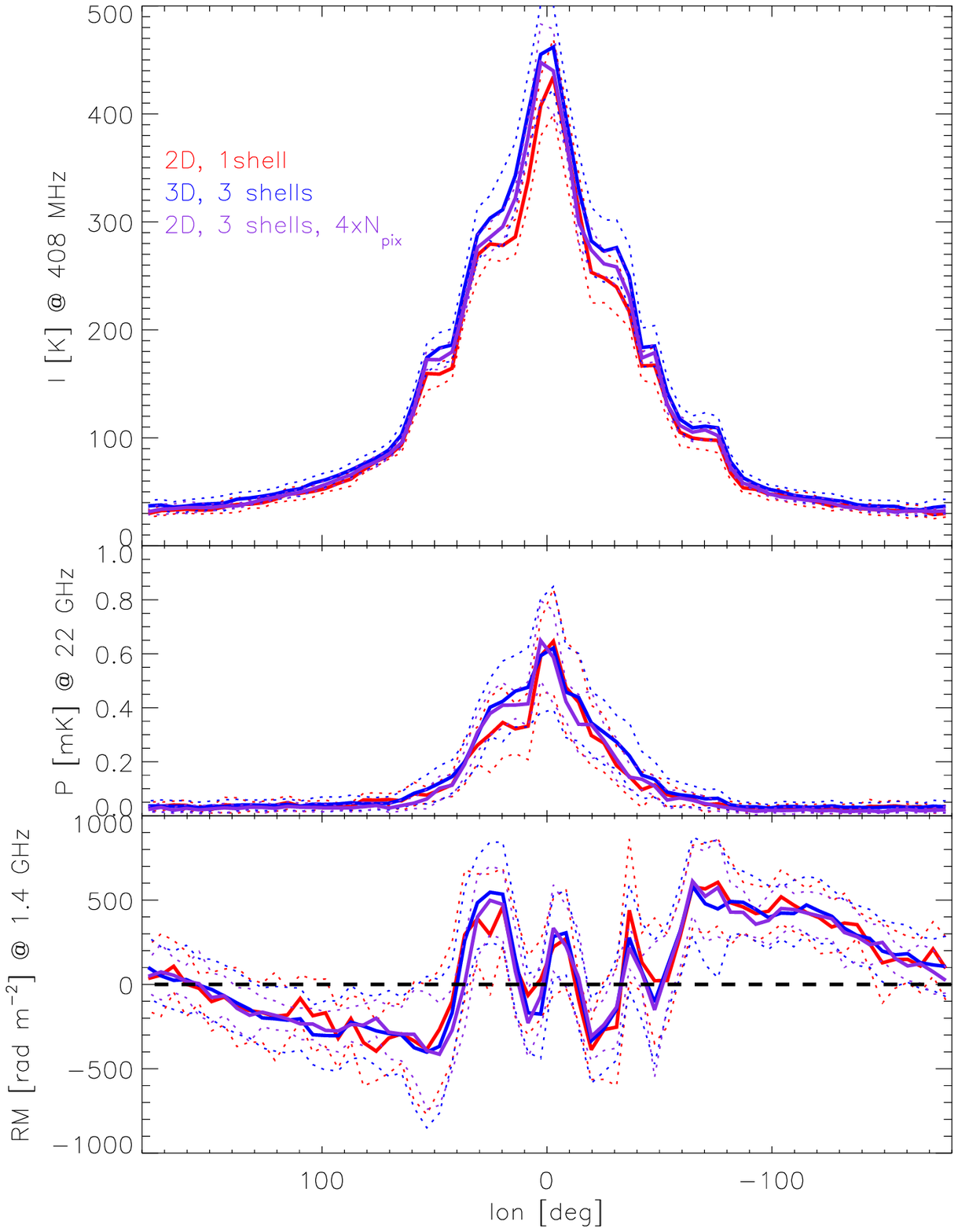}
% originally from
% work/galactic_b/plane_mcmc/test_lowres/test_reso/plots/compare_profiles_2D3D2.ps
\caption{Testing the effects of resolution and dimension of GRF simulation 
as described in \S~\ref{sec:reso}.  The solid lines are the mean
models while the dotted lines indicate the RMS.  The difference
between our 2D analysis and a better but more time-consuming 3D method
is apparent only in the strength of the step-features corresponding to
spiral arm tangents where the small-scale random component is
strongest.  These differences are within the uncertainties but are
systematic and introduce a small bias discussed in \S~\ref{sec:reso}.}
\label{fig:reso}
\end{figure}

Figure \ref{fig:reso} shows a test of the importance of the resolution
level in the simulations along the plane.  Here, we test the
implications of the simplifications we make for computational
efficiency.

The choice of the resolution of the GRF simulation is an important
question.  The GRF normalisation, \eg its RMS over the whole galaxy
box, is a function of the outer scale, the power law index, and the
resolution.  Though the smaller scales are increasingly irrelevant due
to the power law's negative spectral index, there is an effect of not
including all relevant dynamical scales.  We have verified that
quadrupling the resolution appears to have only a small effect on the
results of the simple Galactic plane analysis, as seen in
Fig.~\ref{fig:reso}.

More importantly, we simulate only a 2-dimensional GRF rather than a
full 3-D galaxy, which speeds up the computation by almost a factor of
ten.  Though we are only looking along the Galactic plane, structure
near the plane does enter the beam of each observation, particularly
for volumes further away.  To be fully correct, we would simulate a 3D
galaxy and include contributions from high-resolution pixels just off
the plane by using multiple {\sc hammurabi} shells to simulate a
finite instrument beam.  The differences in the profiles are shown in
Fig.~\ref{fig:reso}.  There is a visible difference in the total
intensity profiles; the blue curve shows the correct profile, while
the red is what we simulate.  The effect of this simplification is to
underestimate the RMS of the random component.  If we perform the MCMC
analysis to fit for the $B_\rmn{RMS}$ and $f_\rmn{ord}$ parameters, we
% From
% work/galactic_b/plane_mcmc/test_lowres/test_reso/test_3d/scripts/idl_journal.pro.bak
find that the latter is unaffected but the former is underestimated by
approximately $7$ per cent.  We consider this acceptable given that our
knowledge of the relative strengths of these components has not
previously been anywhere near that accurate and that there are further
limitations of our analysis discussed in \S~\ref{sec:results}.

%%%%%%%%%%%%%%%%%%%%%%
\subsection{Non-linear parameters}

Some of the parameters are fairly linearly related to the observables.
The arm amplitudes, for example, simply increase the RM in a given
direction, and since the RMs are the best dataset to use to constrain
them, we use only the RM data to determine the $a_n$ parameters (as
seen in Fig.~\ref{fig:mcmc_example}) as well as the angle $\phi_0$.

Parameters related to the synchrotron emission, particularly those
describing the random magnetic field component, are more clearly
non-linear.  Firstly, the emission itself depends on the field to the
power of $\gamma$ as described in Eq.~\ref{eq:synch}.  Secondly, the
observable polarised emission is affected by the random component much
like a 2D random walk, whose mean distance from the starting point is
then a function of the square root of the number of steps, $N$, and
the step size.  In the case of a power law GRF, we have essentially a
superposition of random walks of many step sizes and $N$s.

Furthermore, the variance used in the likelihood calculation
(Eq.~\ref{eq:like}) depends on the particular location in parameter
space.  As described above, this affects the ``noise'' due to the
non-analytic way we compute the model.  We have tested with a set of
simulations to see how well the method works despite these
shortcomings and found that it works surprisingly well.
%  See work/galactic_b/plane_mcmc/test_lowres/sim_set/results.txt
We fixed the other parameters and fit only to $f_\rmn{ord}$ and
$B_\rmn{rms}$, the two most important but perhaps most complicated and
correlated parameters.  Using both the total and polarised intensities
does succeed in constraining these parameters.  In all but one of the
ten simulations, the resulting best-fit positions were within
2-$\sigma$ of the correct values and showing no bias.  An example is
shown in Fig.~\ref{fig:mcmc_example}.

% Originally from
% work/galactic_b/plane_mcmc/test_lowres/sim_set/plots/result_1.ps
%
%\begin{figure}
%\includegraphics[width=\linewidth]{my2dh_all_sim_1.ps}
%\caption{Example of fitting the two most non-linear parameters.  The 
%data fitted were simulated from the values indicated by the dashed
%lines.  The resulting mean and variance are shown by the solid lines.}
%\label{fig:mcmc_example_nl}
%\end{figure}

%%%%%%%%%%%%%%%%%%%%%%%%%%%%%%%%%%%%%%%%%%%%%%%%%%%%%%%%
\section{Results}\label{sec:results}
%%%%%%%%%%%%%%%%%%%%%%%%%%%%%%%%%%%%%%%%%%%%%%%%%%%%%%%%

Our principle aim is to attempt to constrain the ratios of the three
components of the Galactic magnetic field: the coherent component, the
isotropic random component, and the ordered component.  With the three
complementary datasets of total synchrotron emission, polarised
synchrotron emission, and Faraday rotation measures, we have
demonstrated that these three components can be disentangled using
reasonable models of the magnetised ISM.

%%%%%%%%%%%%%%%%%%%%%%
\subsection{RM fits}\label{sec:results_rm}

We begin by constraining the large scale, coherent field.  In other
galaxies, the magnetic field usually shows spiral structure, sometimes
azimuthally symmetric, sometimes not.  We should not assume that the
magnetic spiral follows the DIG spiral; the field strength may or
may not be higher in DIG arms, and even the spiral pitch angles may
not be the same.  But it is not our primary aim in this work to
determine the global morphology of the coherent magnetic field.  We use
the RM data largely to {\em constrain} the amplitude of the emission
due to the coherent field.  Though its morphology may not follow
precisely the spiral model we have fit to the data, it is enough to
limit the amount of synchrotron emission from that component, allowing
us to then examine that from the random and ordered components
independently.

Using the methodology described in \S~\ref{sec:method}, we fit a six
parameter spiral model to the RM data alone to constrain the values of
the following parameters: the orientation angle, $\phi_0$, the
amplitudes $a_0$ to $a_3$ for each of the four arms, and $a_4$ for the
``molecular ring''.  The amplitudes $a_n$ then give the relative
strengths of each feature, which combine with $C_0$ to give the
maximum field strength along the ridge.  The relative amplitudes,
particularly their signs that determine the direction of the field in
each arm (clockwise or counterclockwise when viewed from the north
Galactic pole) determine the features seen in the RM data as in
Figs.~\ref{fig:plane_dataonly} or
\ref{fig:compare_ne2001}.

The parameter fit results are shown in Table~\ref{tab:results}.  This
part of the analysis does not give much that is new compared to what
has been done by
\citet{brown:2007} and others.  We do clearly see the magnetic field
reversals, visible in the RM data, in the general direction of
Scutum-Crux and near the tangent to the proposed ``molecular ring'' at
negative longitudes.  This is visible in comparing Figures
\ref{fig:plane_dataonly} and
\ref{fig:regmodel}.  
%  test_RMonly2 results, using cr0=1, all around 1 in amps, so total
%  reg field at maximum (rho_c=cr0+1) then at most 2.  
We also confirm the field strength peaking at $\approx 2 \,\mu$G in
the arm ridges.  It is interesting to note that, though the field is
free to fit the azimuthal rotation $\phi_0$, the result lies very
close to the DIG spiral arms in the NE2001 model, whether that
model is assumed for the thermal electrons or not.

The value of $a_0$ is not constrained, because as seen in
Fig.~\ref{fig:regmodel}, no sightlines are really tangent to the arm.
It contributes relatively little in the region where we have RM data.

%%%%%%%%%%%%%%%%%%%%%%
\subsection{Synchrotron fits}\label{sec:results_sync}

Our main aim, however, is to attempt to quantify the relative amounts
of coherent, random, and ordered magnetic field components.  As found
by Broadbent at al., the magnetic field must be amplified in the arms.
Since the cosmic rays are likely smooth on these scales, the step-like
features in the synchrotron profile are probably due to the
enhancement of the perpendicular component of the magnetic field in
the arms.  Though the coherent component could be scaled to produce
such structures, it would also require an extremely low
($n_\rmn{e}\approx 10^{-3}\,\rmn{cm}^{-3}$) electron density in order to
be consistent with the RM data.
%See Fig.~\ref{fig:iso_aniso}.

The fit results in Table~\ref{tab:results} for the parameters
$B_\rmn{RMS}$ and $f_\rmn{ord}$ are simply due to the relative
strengths of the total and polarised emission along the plane.  The
resulting best-fit profiles and residuals are shown in
Fig.~\ref{fig:fit_profiles}.  These profiles are not a perfect match
to the total intensity data, but they agree remarkably well in most
places.  The most significant deviation is where there is a strong
thermal feature around longitude -70$\degr$ to -80$\degr$, which is
the region that dominates the high $\chi^2$.  There is also a mismatch
in the profile around longitude 60$\degr$ which bears further
investigating but appears to be related to the Sagittarius arm.  Recall
from \S~\ref{sec:synch} that the innermost region is not included in
the fit.  Further work will be necessary to determine the magnetic
field structure in the Galactic centre region; the model we use,
particularly the central part of the profile shown in
Fig.~\ref{fig:r_prof}, is ill-constrained.

Figure \ref{fig:fit_profiles} shows that the
polarised synchrotron profiles are only fit to first order.  The
$\chi^2$ values are not bad due to the galactic variance, but the
residuals show clear systematic differences where the profile shape is
wrong.  More work will be needed to determine what this means for the
model, particularly the spatial variation of the magnetic field
components.

\subsection{Component ratios}\label{sec:fit_results}

From the results of our fitting the parameters $\phi_0$, $a_n$,
$B_\rmn{rms}$, and $f_\rmn{ord}$, we learn the following:

\begin{enumerate}

\item The rotation measures allow us to constrain the coherent magnetic 
field parameters $a_n$.  As shown in Fig.~\ref{fig:compare_ne2001},
they alone cannot distinguish between the NE2001 spiral model for the
thermal electron density and a simple uniform model.  Not shown on
that figure are the synchrotron profiles, which are effectively
indistinguishable.  The reason is that the synchrotron profile is
apparently dominated by the random and ordered components.  The RMs,
regardless of the thermal electron model, have constrained the coherent
field to an amplitude that is fairly low compared to the synchrotron
profile, assuming that the cosmic ray electron density is roughly
correct.  

Therefore, our results suggest that the coherent field in the arms
peaks around 
%
%  Using b0=cr0=1, a_n around 1.  rho_c is c0+1=2 on ridge
%
$B_0 a_n (C_0+1)= 1-3\,\mu$G in the arms, varying from arm to arm with
an average of 2~$\mu$G.

%
%  try_data2/test_ne2001/chains/all2para* gives
%     B_rms=2.1 and f_ord=1.9 (and cr0=1)
%  while test_scale_ne2001/chains/allbpara gives 2.1 and 1.4
%  so with 
%   (c0+1)*Brms, you get 
%   (1.+1)*2.1=4.2  for unscaled
%   (0.9+1)*2.1=4.0 for scaled
%  where 1.1 (0.9) from averaging a_ns for basic (scaled)
%
\item The isotropic random field component peaks around 
$\left<B_\rmn{iso}^2\right>^{1/2}=(C_0+1) B_\rmn{RMS}=4.2\,\mu$G
(see Eq.~\ref{eq:peaks}).

% Bord= ford*c0*sqrt(2./3.)*brms = 3.26 
%
%  or 1.6*1.*sqrt(2./3.)*2.1=2.7
%
\item The ordered field component peaks slightly lower at $3.3\,\mu$G.

%  All components in uG:
%IDL>  print,[mean(abs(ans))*(c0+1), (c0+1)*brms, ford*c0*sqrt(2./3.)*brms]
%      2.22400      4.20000      3.25782
%
%  Energy density ratios:
%IDL> print,[(mean(abs(ans))*(c0+1))^2, ((c0+1)*brms)^2, (ford*c0*sqrt(2./3.)*brms)^2]/total( [(mean(abs(ans))*(c0+1))^2, ((c0+1)*brms)^2, (ford*c0*sqrt(2./3.)*brms)^2] )
%     0.148983     0.531332     0.319685
%
\item The fractions of the energy density in coherent, random, and 
ordered field components ($\propto \left<B^2\right>$) are then roughly
1:5:3, respectively.  The uncertainties are discussed further in
\S~\ref{sec:disc}.  Note that this model implies a total {\em local}
field strength of roughly 3~$\mu$G, and a mean field strength in the
inner 10~kpc of the Galaxy of 3.5~$\mu$G.

\end{enumerate}

\begin{table}
\begin{center}
\begin{tabular}{|c|c|c|}
%  See 
%
\multicolumn{3}{c}{ Fits to RM data only }\\
\hline
 & {\bf NE2001} & ${\bf n_\rmn{e}=0.1\, \rmn{cm}^{-3} }$ \\
\hline
%
% From
% work/galactic_b/plane_mcmc/try_data2/test_RMonly2{_ne2001}/chains/all6bpara_*
% IDL> my2dh,'../test_RMonly2_ne2001/chains/all6bpara_',nchains=32,doparams=[1,3,4,5,6,7],nbins=15,savefile='my2dh_RMonly2_ne2001_6bpara_idl.dat',charsize=1,'../plots/my2dh_RMonly2_ne2001_6para.eps',smin=500
%
$\phi_0$  & $70 \pm 6.2$ & $70 \pm 7.8$ \\ 
$a_0$  & $1.65 \pm 0.38$ & $-0.15 \pm 0.24$ \\ 
$a_1$  & $0.61 \pm 0.18$ & $0.33 \pm 0.15$ \\ 
$a_2$  & $-1.04 \pm 0.21$ & $-0.87 \pm 0.20$ \\ 
$a_3$  & $1.26 \pm 0.64$ & $0.97 \pm 0.46$ \\ 
$a_4$  & $-1.00 \pm 0.35$ & $-1.33 \pm 0.43$ \\ 
\hline
\multicolumn{3}{c}{ Fits to all data with above NE2001 $n_\rmn{e}$ values fixed }\\
%\hline
 & {\bf constant compression} & {\bf scaled}   \\
\hline
%
% See work/galactic_b/plane_mcmc/try_data2/test_{scale_}ne2001/chains/all2para_
%               constant	scaled
$B_\rmn{RMS}$ &  $ 2.1 \pm 0.03 $ & $ 2.1 \pm 0.04 $ \\
$f_\rmn{ord}$ &  $ 1.9 \pm 0.14 $ & $ 1.5 \pm 0.16 $ \\
\hline
$\chi^2_\rmn{I}/N$  & 3.7 & 3.5 \\
$\chi^2_\rmn{PI}/N$ & 0.8 & 1.0\\
$\chi^2_\rmn{RM}/N$ & 1.8 & 1.6\\

\end{tabular}
\end{center}
\caption{{\em Top:} results of fitting RMs with the simple spiral 
model parameters and comparing constant thermal electron density model
to NE2001.  {\em Bottom:} results of fixing the $\phi_0$ and $a_n$
parameters to the best fit values show on top using NE2001 and all
others to the ``defaults'' listed in Table~\ref{tab:params}, and then
fitting only the $f_\rmn{ord}$ and $B_\rmn{RMS}$ parameters to all
datasets.  As described in \S~\ref{sec:scaling}, the ``constant
compression'' have the same amplitude of random components in each arm
while the ``scaled'' have the factors of $a_n$ applied.}
\label{tab:results}
\end{table}

\begin{figure}
\begin{tabular}{c}
% from work/galactic_b/plane_mcmc/try_data2
%IDL> my2dh,'../test_RMonly2_ne2001/chains/all6bpara_',nchains=32,doparams=[3,4,5,7],savefile='my2dh_RMonly2_ne2001_6bpara_idl.dat',nbins=20,'../plots/my2dh_RMonly2_ne2001_4para.eps',smin=500,plot_labels=['','(a)','(b)','','(c)','(d)'],$
%IDL> charsize=2,doplots=[1,2,4,5]
%
\includegraphics[width=\linewidth]{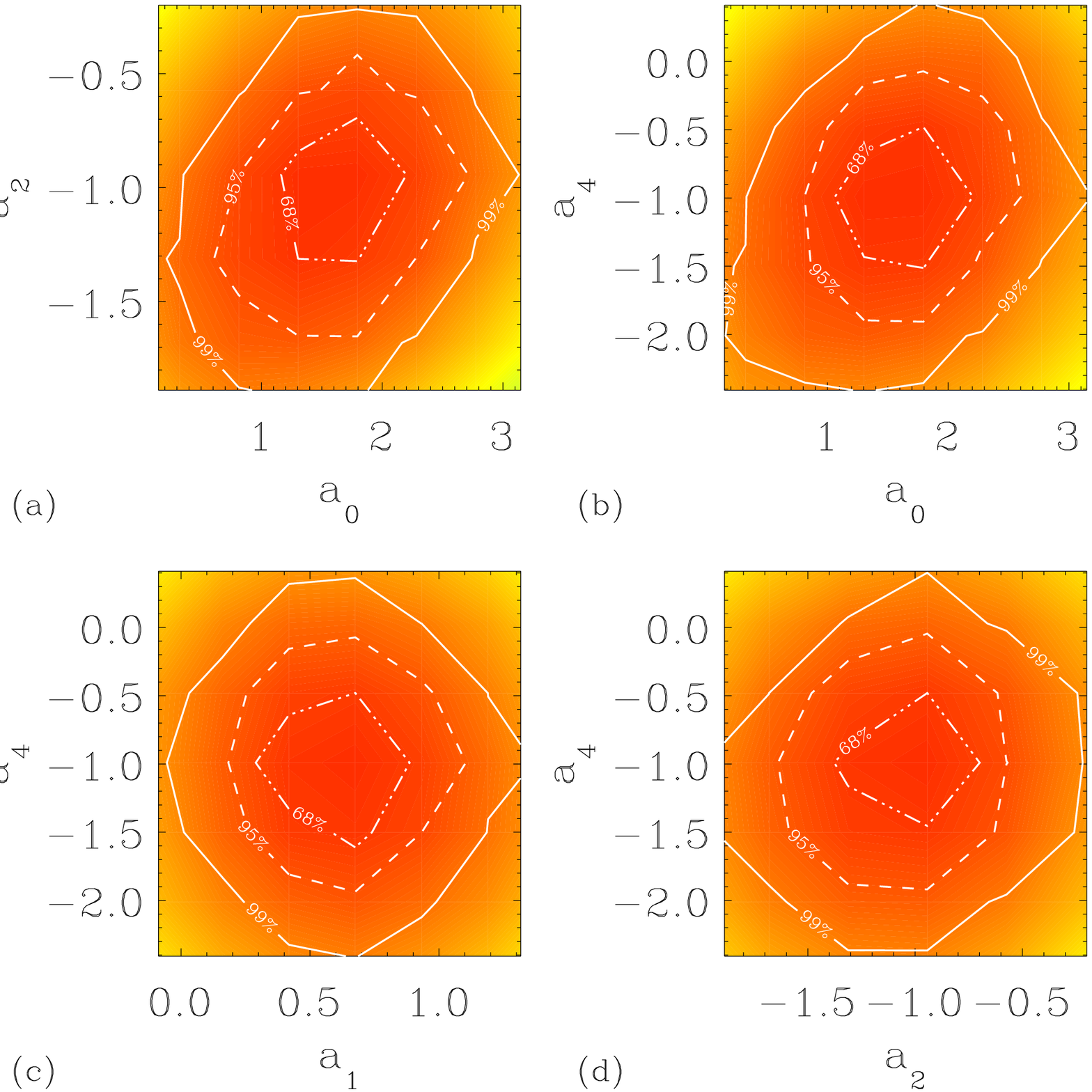}\\
%
%IDL> my2dh,'../test_ne2001/chains/all2para_',nchains=32,doparams=[2,8],nbins=20,smin=100,plot_labels=['(e)'],rfactor=7,smooth=1,'../plots/my2dh_ne2001_2para.eps'
%
\includegraphics[width=0.8\linewidth]{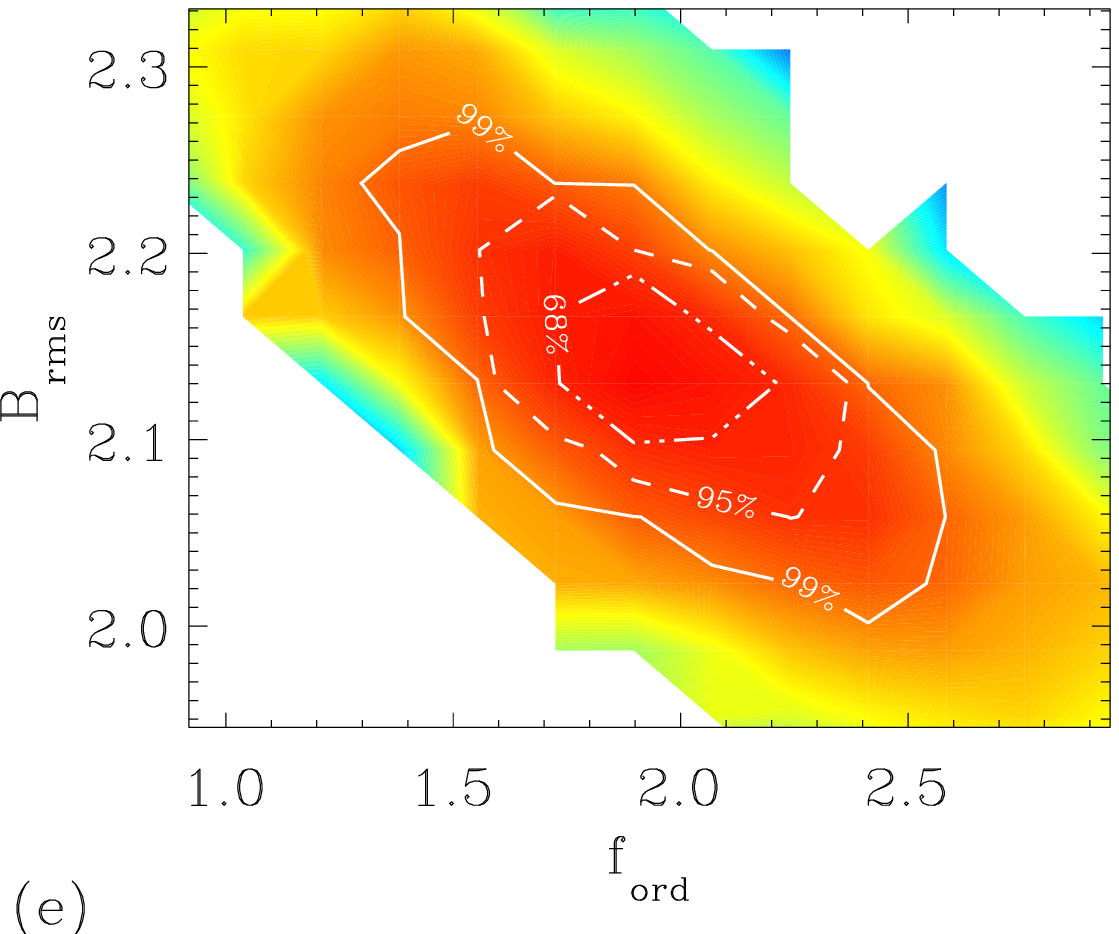}\\
\end{tabular}
\caption{ {\it Top, (a) through (d):}  selected results of fitting 
the coherent field parameters to the RM data only using the NE2001
thermal electron density model.  {\it Bottom, (e):} results of fixing the
coherent field parameters to the values found in the RM analysis, and
then fitting the two parameters controlling the small-scale, irregular
components to all the data.  The means and uncertainties are given in
Table~\ref{tab:results}.  Compare to Fig.~\ref{fig:mcmc_example}.}
\label{fig:results}
\end{figure}

\subsection{Arm strength}\label{sec:scaling}

In the basic analysis, the arm strengths, $a_n$, determine the coherent
field strength in each arm (as well as direction) but do not affect
the random or ordered components.  For simplicity, these components
are simply scaled by $\rho_c$ as described in Table~\ref{tab:params},
which is only a function of the distance from the ridge.  We can also,
however, adjust the strength of these components by the same factors
by multiplying $C_0$ by $|a_n|$.  This makes the contrast in each arm
proportional to the coherent field strength of that arm rather than
constant.  (It also implies that the ratio of the ordered to isotropic
random fields changes from arm to arm.)  

\begin{figure}
\includegraphics[width=\linewidth]{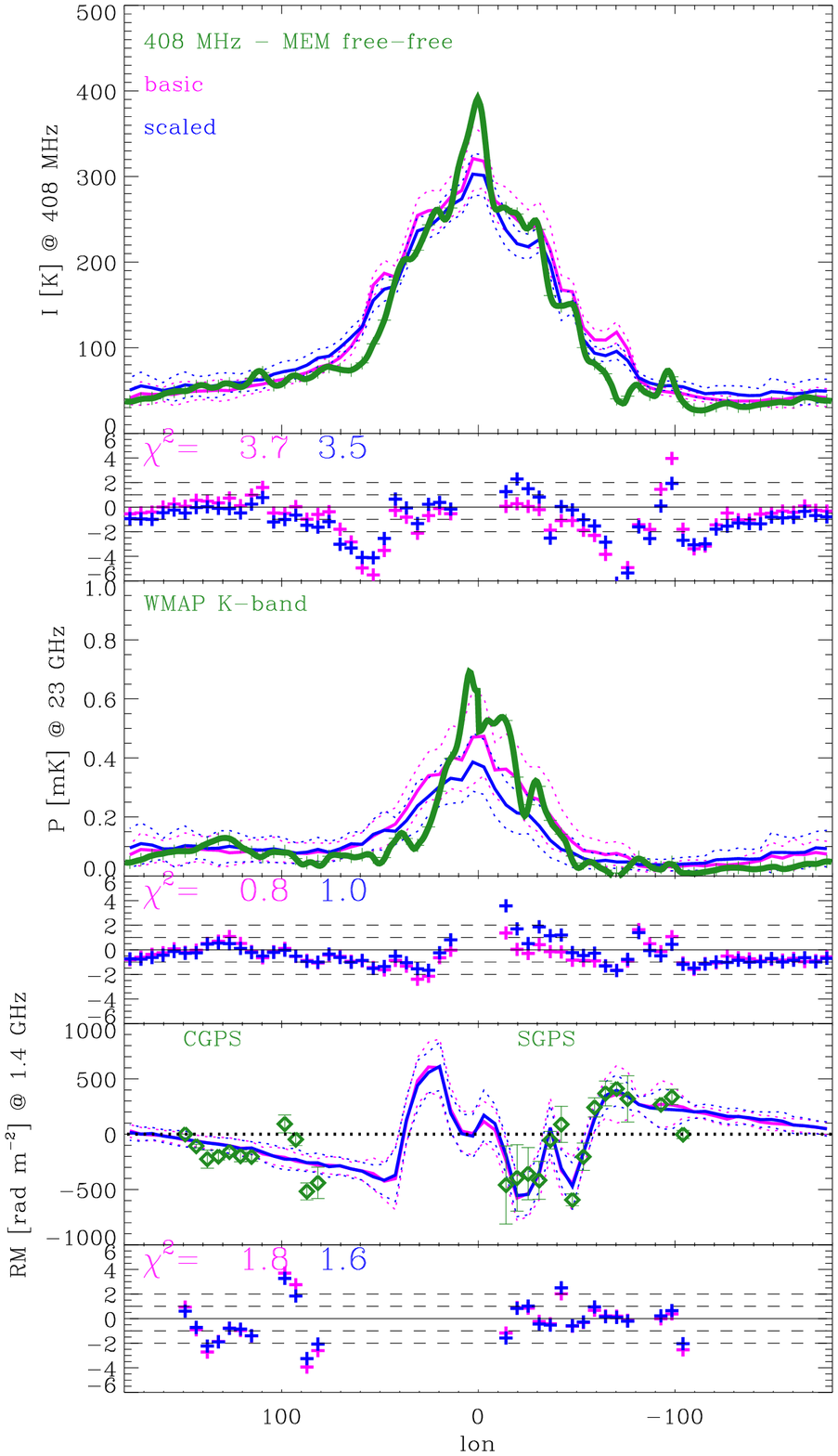}
\caption{Best fit models from MCMC analysis as described in 
\S~\ref{sec:results_rm} and \ref{sec:results_sync}.  The three observables 
are shown from the top as synchrotron total intensity at 408~MHz,
synchrotron polarised intensity at 23~GHz, and rotation measure.  The
solid blue and magenta lines are the mean models while the dotted
lines indicate the RMS, or galactic variance, and in thick green are 
the data.  For each observable, beneath the plotted profile are the
residuals, \ie $(data-model)/error$, compared to dashed lines at
$\pm$1 and $\pm 2\,\sigma$.
In magenta is the result where all arms have equal random
and ordered compression, while in blue is the result where those
components are also scaled by $a_n$ as described in
\S~\ref{sec:scaling}.}
\label{fig:fit_profiles}
\end{figure}

A comparison of the best-fit profiles is shown in
Fig.~\ref{fig:fit_profiles}.  The physical motivation for scaling each
arm strength independently is unclear, but the result including the
extra scaling appears to fit the synchrotron profile slightly better
in some places and slightly worse in others.

%
%  work/galactic_b/plane_mcmc/try_data2
% IDL> plot_resids_vert,['0','0'],roots=['../plotting/result_2para_basic_ne2001/out/','../plotting/result_2para_scaled_ne2001/out/'],labels=['basic','scaled'],psfile='../plots/resids_2para_ne2001.eps'
%

\subsection{Arm/inter-arm Contrast}\label{sec:contrast}

In the basic analysis presented above, we have fixed the value of the
parameter $C_0$ that defines the contrast between the arm ridges and
inter-arm regions.  We can instead allow this parameter to vary as
well.  The combination of $C_0$ and $B_\rmn{RMS}$ controls the
relative strength of the arm versus the inter-arm region for the
random component (as $C_0$ and $B_0$ or $a_n$ does for the coherent
component).  

\begin{figure*}
\includegraphics[width=0.8\linewidth]{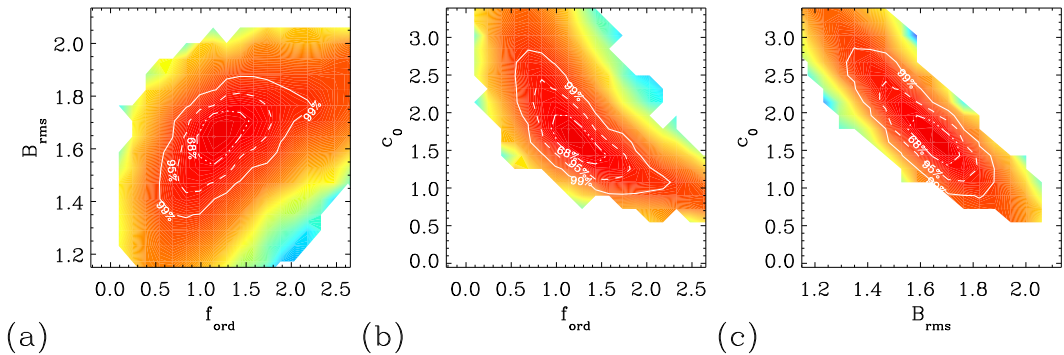}
\caption{MCMC results showing the degeneracies described in 
\S~\ref{sec:contrast} among the three 
parameters: $B_\rmn{RMS}$, $f_\rmn{ord}$, and $C_0$. }
\label{fig:contrast}
\end{figure*}

These three parameters, however, are somewhat degenerate as shown in
Fig.~\ref{fig:contrast}.  The data do not easily distinguish between a
stronger $B_\rmn{RMS}$ with a weaker contrast ($C_0$) or {\it vice
versa}.  This is an example of one of many degeneracies in the
parameters listed in Table~\ref{tab:params}.  But our basic result
comparing the strengths of the components {\it in the arms} is not
significantly affected.

%
% work/galactic_b/plane_mcmc/try_data1/plots/my2dh_all3d.ps
%  from                                chains/all3d*txt
%

\subsection{Discussion}\label{sec:disc}

We have used the three complementary datasets of total synchrotron
emission, polarised synchrotron emission, and rotation measure to
study the three components of the Galactic magnetic field, namely the
coherent, random, and ordered fields.  We find that we can fit the
profiles of these datasets along the plane with a model where the
components peak in magnetic arm ridges and where the coherent component
contributes roughly 10 per cent to the energy density ($\left<B^2\right>$),
while the random and ordered components contribute roughly 50 and
40 per cent, respectively.  

This is a first attempt to constrain models with many parameters and
degeneracies.  If the estimates for the CRE and thermal electron
densities are even approximately right, our method gives a fairly
robust constraint on the {\em relative} strengths of the three field
components.  We should, however, quantify the impact of such
assumptions and their uncertainties:

\begin{itemize}

\item There is a degeneracy between the thermal 
electron density in the arms and the amplitude of the coherent magnetic
field.  A factor of two uncertainty in the mean electron density in
the arms implies a factor of two uncertainty in the coherent field
strength.
%
%  btot=(cr0+1)*(breg+brms) + cr0*ford*sqrt(2./3)*brms
%IDL> print,(c0+1)*(mean(abs(ans))+brms) + c0*ford*sqrt(2./3.)*brms
%      9.68182
% 
%  If you drop n_e by a factor of 2 and then increase 
%  Breg by a factor of 2, then solve for Brms, you get 
%IDL> print, (btot-(c0+1)*mean(abs(ans))*2.)/( c0+1+c0*ford*sqrt(2./3.) )
%      1.47376
% which is 70% of the original 2.1
%
% B field contributing to P only reg and ordered.  Solve for ford:
%IDL> bptot=(c0+1)*mean(abs(ans)) + c0*ford*sqrt(2./3.)*brms
%IDL> print,(bptot-(c0+1)*mean(abs(ans))*2)/( c0*sqrt(2./3.)*1.47 )
%     0.861338
% but B_ord is ford*c0*Brms*sqrt(2/3), so it ends up 1.03 uG, or half
%
% so ratios are
%IDL> print,[(mean(abs(ans))*(c0+1)*2)^2, ((c0+1)*brms*0.7)^2, (ford*0.86*c0*sqrt(2./3.)*brms*0.7)^2]/total([(mean(abs(ans))*(c0+1)*2)^2, ((c0+1)*brms*0.7)^2, (ford*0.86*c0*sqrt(2./3.)*brms*0.7)^2])
%     0.613011     0.267814     0.119175
%
If the electron density is a factor of two smaller than our model for
a given arm, the coherent field is then a factor of two larger to
reproduce the same RM profile.  This then requires a drop in
$B_\rmn{rms}$ to reproduce the same synchrotron total intensity
profiles.  The ratios of energy densities of the components
(coherent:random:ordered) would go from roughly 1:5:3 to 6:3:1.

\item Similarly, in using the NE2001 model for a smoothly
distributed, average electron density, we have assumed a relatively
homogeneous DIG, while clearly the density of free electrons varies
many orders of magnitude from HII regions down to molecular clouds.
\citet{berkhuijsen:2006} and others have
studied it's clumpiness as described by a filling factor.  The LOS
filling factor is often defined as $f\equiv
\left<n_e^2\right>/\bar{n}_c^2$, where the brackets denote the average
along the entire LOS while $\bar{n}_c$ is the average value in the
individual clumps; it then also represents the fraction of the LOS
that falls within the clumps.  This can also be defined as
$f=\left<n_e\right>^2/\left<n_e^2\right>=DM^2/(EM\,D)$ (where $D$ is
the total LOS distance, $EM$ and $DM$ are the emission and dispersion 
measures, respectively).  We note that the factor derived from
comparing EM and DM oversimplifies the fact that EM and DM essentially
trace different physical components of the ISM.

The NE2001 is largely based on
DM, which is linear in $n_e$ and path length, and so any factor of
$f\neq 1$ cancels.  If the fluctuations in $n_e$ are uncorrelated with
the magnetic field, then RM is similarly unaffected on average.  It is
unclear how correlated the field is with the electron density,
however, so the impact of this approximation is difficult to judge.
If the field fluctuations along the LOS are equally likely to be
toward or away from the observer in high density clumps, then even if
the strength of the fluctuations is correlated with $n_e$, this will
not affect the predicted mean RM profile, only its variance.  If,
however, the correlated field fluctuations retain a coherent
direction, then the observed RM would be larger than what we have
modelled.  Properly taking this into account would then require a
weaker field to reproduce the same RM data.  In future, we may be able
to plug in the results of magneto-hydrodynamic (MHD) turbulence
simulations into the tools we have described to test this explicitly.

\item There is a degeneracy between the magnetic field strength 
and the cosmic ray electron density.  We assume a value of
$J_\rmn{CRE,\oplus}=0.25\,\left(\rmn{GeV\,m^2\,s\,sr}\right)^{-1}$ at
the position of the Sun.  If the uncertainty in this number is a
factor of two (which it may well be; see, \eg \citealt{strong:2004})
and assuming $B_\rmn{coh}$ is constrained by the RM data, then we can
solve for the other components.
%
%  I=ncre*Btot^2=ncre*Bsq
%   
%  Bsq=((c0+1)*(Breg+brms)+ c0*ford*sqrt(2./3)*brms)^2
%
%  If B^2 must be double/half that (because ncre is halved/doubled), then 
%IDL> print,(sqrt(2*bsq)-(c0+1)*mean(abs(ans)))/( c0+1+c0*ford*sqrt(2./3.) )
%      3.22925
%IDL> print,(sqrt(0.5*bsq)-(c0+1)*mean(abs(ans)))/( c0+1+c0*ford*sqrt(2./3.) )
%      1.30150
%
%the $B_\rmn{rms}$ is uncertain by $\approx 40-50\%$.  
%
%  Likewise,
%
%  PI=ncre*Bptot^2=ncre*( (c0+1)*Breg + c0*ford*sqrt(2./3.)*brms)^2
%  so ford is
%
% IDL> print,(sqrt(2.*bpsq)-(c0+1)*mean(abs(ans)))/(c0*3.2*sqrt(2./3.)) & print,(sqrt(0.5*bpsq)-(c0+1)*mean(abs(ans)))/(c0*1.3*sqrt(2./3.))
%      2.11592
%      1.55659
%
% i.e. varying by 10-20%
% So componts that were 2.2,4.2,3.2 uG are then 2.2,6.4,5.5 or 2.2,2.6,1.7
% and then the ratios from 1:5:3 to 1:5:4 (CRE halved, B^2 doubled) to 3:5:2
The ratios of energy densities would then vary from  0.6:5:4 (CREs
halved) to 3:5:2 (CREs doubled).

The recent results from the Fermi $\gamma$-ray telescope
\citep{abdo:2009} seem to indicate a density that is somewhat lower
than what we use from \citet{strong:2007} based on EGRET data.  A
lower CRE density implies a larger random magnetic field component to
reproduce the observed synchrotron emission.  (The work of
\citet{sun:2008} use an even higher density of 
$J_\rmn{CRE,\oplus}=0.4\,\left(\rmn{GeV\,m^2\,s\,sr}\right)^{-1}$,
which partly explains how they can reproduce roughly the observed
amount of emission with only an isotropic random component and no
ordered component.)

\item We have assumed a spatially smooth cosmic-ray electron 
distribution with a simple power-law power spectrum that is the same
throughout the Galaxy and has index $p=3$.  A spatial variation in
the spectrum or deviations of the spectrum from the assumed
power law would also affect our analysis.

As an example, we can imagine that the CRE spatial and spectral
distribution varied between spiral arm and inter-arm regions due to
the higher density of CRE acceleration sites (\eg supernova remnants)
in the arms (see, \eg \citealt{case:1996}).  This would have a similar
effect as the global uncertainty discussed above.

Furthermore, in assuming this power law distribution of electrons, we
assume that the spectrum of synchrotron emission in brightness temperature
also follows a power law with $\beta=-(p+3)/2=-3$.  Analyses such as
\citet{platania:1998} or \citet{giardino:2002} show that this index in fact 
varies across the sky.  Determining this index on the plane over a
broad frequency range is complicated by the variety of emission
mechanisms that contribute to total intensity and by the Faraday
effects that contribute to polarised intensity at low frequencies.  If
our assumption of $\beta=-3$ is too steep, our model will be
underpredicting the polarised emission at 23~GHz from the coherent
magnetic field component, for example.  This is difficult to quantify,
however, since both the synchrotron emissivity (as $B^\frac{p+1}{2}$)
and the polarisation fraction ($\Pi=\frac{p+1}{p+7/3}$) depend on $p$.

\item There are uncertainties in the zero-levels of the synchrotron
datasets we used.  These uncertainties are relatively small (\eg a few
K for the 408~MHz map), usually much less than the galactic variance,
and will be most significant where the synchrotron emission profile is
lowest toward the Galactic anticentre.  For our analysis of the
magnetic field component ratios in the spiral arm peaks, this
uncertainty is certainly not significant.  It will, however, affect
the parameters describing the strength of the field and the density of
CREs in the outer region of the galaxy.  As we have discussed above,
these two distributions are degenerate and not constrained by our
analysis.  In future work, when we add dust emission to the analysis
to break the degeneracy, we will also have to address the zero-levels
more accurately.  For this work, however, we consider the impact of
the uncertainty minimal.

\item We have assumed that the large-scale coherent field and the 
small-scale field components are all strongest in the same regions,
the magnetic spiral arm ridges.  There is some evidence in one galaxy
that the coherent field, in contrast to the random field, may be
stronger in the inter-arm regions \citep{beck:1996}.  Further
work will be needed to determine whether we can distinguish these
possibilities.

\end{itemize}

Our results are the ratios of the field components' energy densities
that we measure as roughly 1:5:4 (coherent:random:ordered).  The above
uncertainties most significantly affect the fraction of magnetic field
energy in the coherent component.  The rough estimates above then
imply that the coherent component could make up as little as 5 percent
of the total energy in the field, or as much as 30 per cent given an
uncertainty of a factor of 2 in the CRE density, or as much as 60 per
cent if the thermal electrons are half the assumed value.  The latter
is certainly possible for a small region, but on the whole, pulsar
distance measures averaged over the galaxy are unlikely to allow such
a large discrepancy.  As mentioned above, Fermi estimates of the CRE
density are even lower than what we assumed, implying even larger
random and ordered components.

There have been other efforts to estimate the amount of turbulence in
the magnetic field.  But one must be careful to understand what it is
that is being measured, as these works generally distinguish only
between a ``regular'' field and a ``turbulent'' field.  Depending on the
observable, what we define as the ordered field may count as either.
\citet{rand:1989} and \citet{ohno:1993}, for example, both look at 
RM data and measure a turbulent magnetic field component of
$4-6\,\mu$G that corresponds to the sum of our ordered and isotropic
random components.  \citet{schnitzeler:2007} give an upper limit of
$B_\rmn{turb}/B_{\rmn{reg},\perp}\approx 2$ based on synchrotron
emission.  Their ``regular'' component is, when only looking at
synchrotron emission, the total of our coherent and ordered
components.
%That measure refers the total synchrotron
%emission toward the anti-centre, which would include the Perseus arm
%as well as the inter-arm regions to either side.
\citet{haverkorn:2004c}, by contrast, find a random component smaller than 
the ``regular'' component, though this is in two relatively small fields
that may be probing smaller-scale structure only.
\citet{deschenes:2008} find $B_\rmn{turb}/B_\rmn{reg,\perp}=0.57$ using 
synchrotron polarisation in \wmap data, so again, the ``regular''
component includes what we define as ordered.  Each of these studies
uses different observables, different regions of the Galaxy, and
different assumptions.  It is thus difficult to compare their results
with each other or with our estimate, but with the exception of the
Haverkorn \etal result, they are roughly consistent.

What we can conclude, however, is that the coherent field is likely a
relatively small fraction of the total and that a model including only
an isotropic random component and a coherent field is incomplete.  To
reproduce all of the observables, particularly the polarised emission,
requires an ordered field component as well.

%%%%%%%%%%%%%%%%%%%%%%%%%%%%%%%%%%%%%%%%%%%%%%%%%%%%%%%%%%%%%%%%
\section{Conclusions}

We have outlined a method to simulate observables such as synchrotron
emission and Faraday rotation measure based on parametrised models of
the physical components of the Galaxy and to use these to study the
complicated parameter space in an MCMC analysis.  In particular, we
describe how to deal with the galactic variance due to the fact that
one of the principle physical components is stochastic.  These tools
can be used to model the global properties of the Galaxy's magnetic
field in 3D or to study the turbulent component of the magneto-ionic
medium in detail without the need for as many simplifying assumptions.  

To demonstrate the method's utility, we have used the three
complementary datasets of total synchrotron emission, polarised
synchrotron emission, and rotation measure to try to constrain the
relative contributions of the coherent, random, and ordered components
of the magnetic field.  We find the relative energy densities of these
components to be roughly 1:5:3, respectively, at their peaks in a
magnetic spiral arm model.  If our assumed CRE and thermal electron
distributions are roughly correct, then this implies arm field
strengths of 2, 4, and 3~$\mu$G, respectively.  This particular
analysis is limited by several simplifying assumptions, but we
consider the result a first step and proof-of-concept showing
how to tackle this complicated problem.  

In future work, we will address some of the limitations and
degeneracies discussed above.  More importantly, we look forward to
the prospect of additional data.  In the next few years, we will have
the C-Band All Sky Survey (C-BASS)\footnote{{\tt
http://www.astro.caltech.edu/cbass/}} at 5~GHz giving a much more
sensitive measurement of the polarised synchrotron emission in the
radio bands.  These data will help us to study the spectral variation
of the synchrotron emission, and in turn the distribution of cosmic
ray electrons.

The Planck satellite \citep{planck:2006} will also provide a higher
sensitivity polarised synchrotron sky map in the microwave bands when
combined with continuing observations by \wmap.  Its many bands will
also contribute, along with C-BASS and
\wmap, to a much more accurate separation of the thermal emission than
is currently possible.  This will improve our fitting of the step
features in the synchrotron emission along the plane where the thermal
emission is a problem.  Furthermore, the High Frequency Instrument on
Planck will give us a much better map of the polarised dust emission,
a completely independent tracer of the magnetic fields that we have
not used in this work.

The GALFACTS \footnote{{\tt http://www.ucalgary.ca/ras/GALFACTS/}}
survey at Arecibo Observatory will give us the vital coverage of the
northern side of the Galactic centre in rotation measures (see
Fig.~\ref{fig:plane_dataonly}).  This will be crucial in distinguishing the
various models of the magnetic field reversals currently so
controversial.  And that in turn will inform theories such as dynamo
models for the origin, amplification, and evolution of the coherent
field.

The Australian Square Kilometre Array Pathfinder (ASKAP)
\footnote{{\tt http://www.atnf.csiro.au/projects/askap/}} will significantly 
improve our southern sky coverage of rotation measures.  Even better
will be the Square Kilometre Array\footnote{{\tt
http://www.skatelescope.org/}} itself.  \citet{johnston:2008} and
\citet{gaensler:2004} describe how these will significantly advance 
magnetic field studies.  Not only will they inform our knowledge of
our own galaxy but allow us to make similar observations of external
galaxies using background polarised sources.

%%%%%%%%%%%%%%%%%%%%%%%%%%%%%%%%%%%%%%%%%%%%%%%%%%%%%%
\section*{Acknowledgements}
%%%%%%%%%%%%%%%%%%%%%%%%%%%%%%%%%%%%%%%%%%%%%%%%%%%%%%

The authors would like to thank A. Strong and C. Dickinson for useful
discussions.  We acknowledge use of the HEALPix software
\citep{healpix} for some of the results in this work.  We acknowledge
the use of the Legacy Archive for Microwave Background Data Analysis
(LAMBDA). Support for LAMBDA is provided by the NASA Office of Space
Science.

%%%%%%%%%%%%%%%%%%%%%%%%%%%%%%%%%%%%%%%%%%%%%%%%%%%%%%
\bibliographystyle{mn2e}
\bibliography{references}
%%%%%%%%%%%%%%%%%%%%%%%%%%%%%%%%%%%%%%%%%%%%%%%%%%%%%%

%%%%%%%%%%%%%%%%%%%%%%%%%%%%%%%%%%%%%%%%%%%%%%%%%%%%%%

%%%%%%%%%%%%%%%%%%%%%%%%%%%%%%%%%%%%%%%%%%%%%%%%%%%%%%

\bsp

\label{lastpage}

\end{document}